\documentclass[aps,prd,showpacs,onecolumn,preprintnumbers,notitlepage,
nofootinbib,tightenlines,10pt]{revtex4-1}
\usepackage{amsmath,bm}
\usepackage{times}
\usepackage{amsfonts}
\usepackage{amssymb}
\usepackage{braket}
\usepackage{color,graphicx}
\usepackage{epstopdf}
\usepackage{latexsym}
\usepackage[dvipsnames]{xcolor}
\usepackage{slashed}
\usepackage{CJK,upgreek,fancyhdr}
\usepackage{float}
\usepackage{multirow}
\usepackage{hyperref}
\hypersetup{colorlinks,citecolor=nicegreen,linkcolor=nicered,urlcolor=RoyalBlue}
\usepackage{enumitem}
\usepackage{natbib}
\usepackage{relsize}
\usepackage[left=2.5cm,right=2.5cm,top=2.0cm,bottom=2.5cm]{geometry}
\newcommand{\beq}{\begin{eqnarray}}
\newcommand{\eeq}{\end{eqnarray}}
\newcommand{\non}{\nonumber\\ }

\newcommand{\psl}{ P \hspace{-2.8truemm}/ }
\newcommand{\nsl}{ n \hspace{-2.2truemm}/ }
\newcommand{\vsl}{ v \hspace{-2.2truemm}/ }
\newcommand{\epsl}{\epsilon \hspace{-1.8truemm}/\,  }

\def\lsim{ {\ \lower-1.2pt\vbox{\hbox{\rlap{$<$}\lower6pt\vbox{\hbox{$\sim$}
}}}\ } }
\def\gsim{ {\ \lower-1.2pt\vbox{\hbox{\rlap{$>$}\lower6pt\vbox{\hbox{$\sim$}
}}}\ } }

\def \plb{  Phys. Lett. B }

\def \prd{  Phys. Rev. D }

\def \jhep{ J. High Energy Phys.  }

\definecolor{Red}{rgb}{1.,0.,0.}
\definecolor{Blue}{rgb}{0.,0.,1.}
\definecolor{RoyalBlue}{rgb}{0.0, 0.14, 0.4}
\definecolor{nicered}{rgb}{0.7,0.1,0.2}
\definecolor{nicegreen}{rgb}{0.1,0.4,0.2}

\newcommand{\Rlue}[1]{{\color{RoyalBlue}{#1}}}
\def\orcid#1{\kern .08em\href{https://orcid.org/#1}{\includegraphics[keepaspectratio,width=0.75em]{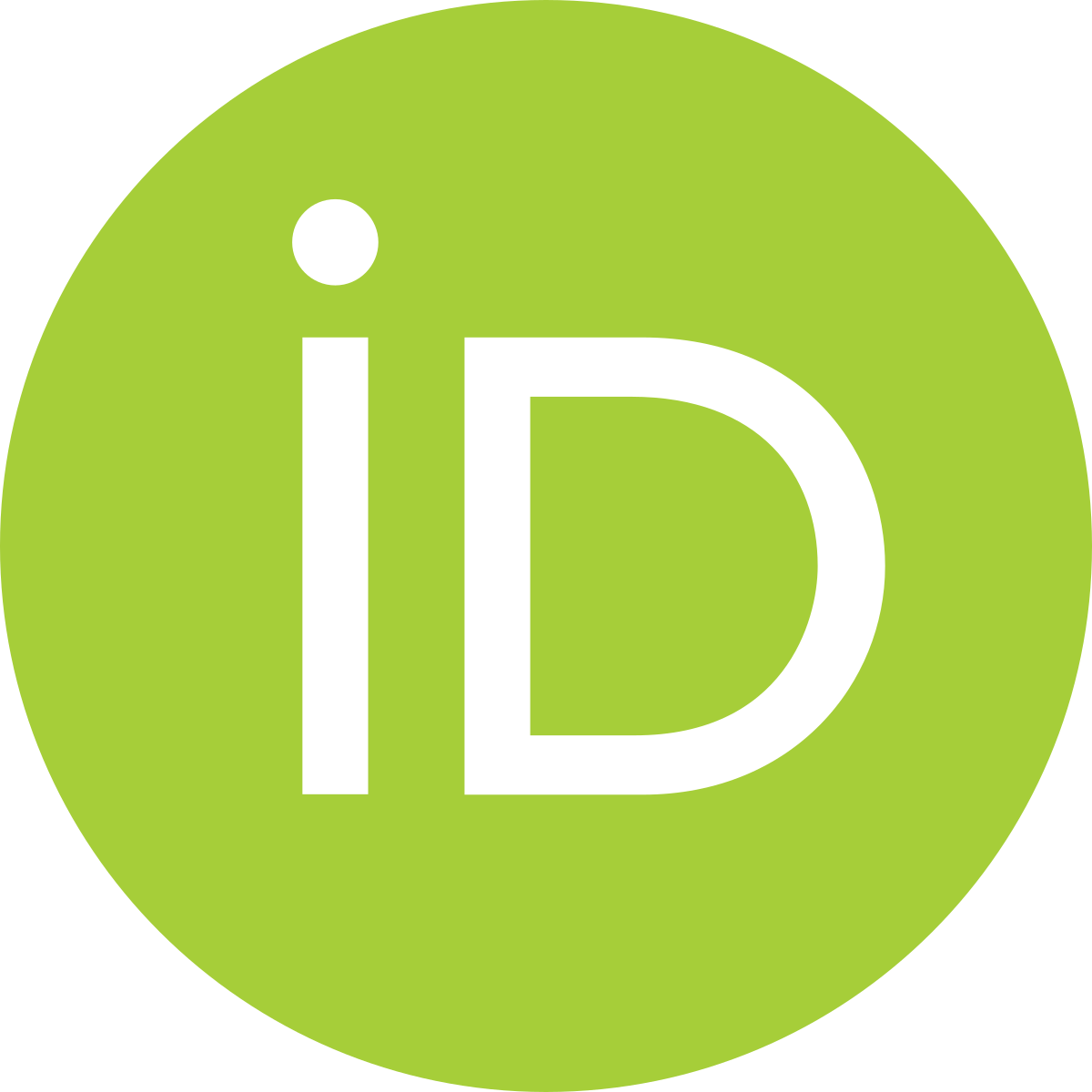}}}

\begin{document}
\begin{CJK*}{GB}{gbsn}
\title{\boldmath
$B_c$-meson decays into $J/\psi$ plus a light meson in the improved perturbative QCD formalism}
%%%==================================================================
\author{Xin~Liu\orcid{0000-0001-9419-7462}}
%\email
%[ Electronic address: ]
%{liuxin@jsnu.edu.cn}
\affiliation{Department of Physics,
Jiangsu Normal University, Xuzhou 221116, People's Republic of China}

%%%%%%%%%%%%%%%%%%%%%%%%%%%%%%%%%%%%%%%%%%%%%%%%%%%%%%%%%%%%%%%%%%%%%

\date{\today{}}

%%%%%%%%%%%%%%%%%%%%%%%%%%%%%%%%%%%%%%%%%%%%%%%%%%%%%%%%%%%%%%%%%%
\begin{abstract}

In the wake of measurements on $B_c^+ \to J/\psi K^+$,
$B_c^+ \to J/\psi \pi^+\pi^-\pi^+$, and $B_c^+ \to J/\psi
K^+ K^-\pi^+$ at Large Hadron Collider experiments,
we propose to study the decays $B_c^+ \to J/\psi M^+$
comprehensively, with $M$ being the light charged pseudoscalar ($P$),
vector ($V$), scalar ($S$), axial-vector ($A$), and tensor ($T$) mesons,
within the improved perturbative QCD (iPQCD) formalism at leading order
in the Standard Model. The theoretical predictions for experimental
observables such as branching fractions, relative ratios, and
longitudinal polarization fractions in the iPQCD formalism
await near future examinations relying on the upgraded Large Hadron Collider,
even the forthcoming Circular Electron-Positron Collider.
We emphasize that the investigations on the factorizable-emission-suppressed
or -forbidden decays like $B_c^+ \to J/\psi S^+$,
$B_c^+ \to J/\psi A^+_{1^1\!P_1}$, and $B_c^+ \to J/\psi T^+$, should go definitely
beyond naive factorization to explore the rich dynamics, which could, in turn,
further help understand the QCD nature of $B_c$ meson, as well as that of related hadrons.
The future confirmations on those predictions about the relative ratios
between the branching fractions of $B_c^+ \to J/\psi b_1(1235)^+ (a_0(980)^+,
a_0(1450)^+, a_2(1320)^+)$ and  $B_c^+ \to J/\psi \pi^+$ could further
examine the reliability of this iPQCD formalism.
Because of containing only tree-level $\bar b \to \bar c$ transitions,
the {\it CP} asymmetries in the $B_c^+ \to J/\psi M^+$ decays exhibit naturally zero.

\end{abstract}

%%%%%%%%%%%%%%%%%%%%%%%%%%%%%%%%%%%%%%%%%%%%%%%%%%%%%%%%%%%%%%

\preprint{\footnotesize  JSNU-PHY-HEP-01/23}
\maketitle

%%
%%%%%
%\tableofcontents{}
%%%%%
%%

\newpage
%
%%%
%%%%%%%%%%%%%%%%% I. INTRODUCTION %%%%%%%%%%%%%%%%%%%%%%%%%%%%%%%%
%%%
%

%\section{Introduction}\label{sec:Intro}

In 1998, the first discovery of $B_c$ meson at Tevatron~\cite{CDF:1998axz,CDF:1998ihx}
proclaimed the beginning of its experimental studies.
After that, its properties, e.g., lifetime and mass, were
measured combined through semileptonic decay $B_c^\pm \to J/\psi
\ell^\pm \nu_\ell$~\cite{CDF:2006kbk,D0:2008thm} and nonleptonic
decay $B_c^\pm \to J/\psi \pi^\pm$, the so-called ``golden channel" in $B_c$-meson
decays~\cite{CDF:2007umr,D0:2008bqs}. Ever since the running of Large Hadron Collider (LHC)
in 2009, the attention has always been be payed from the community of heavy flavor
physics on the measurements of $B_c$-meson decays. The underlying
reason is that a $B_c$ meson is the ground state of a unique meson
family containing two different kinds of heavy flavor, namely, $b$
and $c$, in the Standard Model~\cite{Brambilla:2004wf,Brambilla:2010cs}
and it is the only meson whose decays of both constituents compete with
each other. The $B_c$-meson decays contain rich dynamics in the perturbative regimes,
besides the nonperturbative nature. What is more, its decays might shed
light on possible new physics beyond the Standard Model (for
example, see very recent literature~\cite{Li:2018lxi,Huang:2018nnq,Cheung:2020sbq,Elahi:2021jia}
and references therein).

The LHC experiments have measured several nonleptonic decay channels of $B_c$ meson
\cite{ParticleDataGroup:2020ssz,HFLAV:2022pwe}, however, unfortunately, the exact values
of individual branching fractions (${\cal B}$) for those observed decays are not available
yet because of experimentally complicated background with proton-proton collisions at LHC.
Among them, the $B_c$-meson decays into $J/\psi$ plus a light meson such as
$B_c^+ \to J/\psi a_1(1260)^+$ (In the following context, we will describe
it as $a_1^+$ for convenience, unless otherwise stated.) and $B_c^+ \to J/\psi K^+$
were observed through the relative ratios of branching fractions between the related $B_c^+$ decays. Explicitly,
\begin{itemize}
    \item
In 2012, the decay $B_c^+ \to J/\psi \pi^+ \pi^- \pi^+$ was reported
for the first time by the Large Hadron Collider-beauty (LHCb) Collaboration~\cite{LHCb:2012ag}.
The ratio between the branching fractions of $B_c^+ \to J/\psi \pi^+ \pi^- \pi^+$ and
$B_c^+ \to J/\psi \pi^+$ was measured to be
\beq
R_{3\pi/\pi}^{\rm Exp}
&\equiv&
\frac{{\cal B}(B_c^+ \to J/\psi \pi^+ \pi^- \pi^+)}{{\cal B}(B_c^+ \to J/\psi \pi^+)}
= 2.41 \pm 0.45\;,
\label{eq:ex-r3pp-2012-LHCb}
\eeq
where ``the background-subtracted distribution of the $M(\pi^+ \pi^- \pi^+)$ mass for
the $B_c^+ \to J/\psi \pi^+ \pi^- \pi^+$ data exhibits an $a_1$ peak"~\cite{LHCb:2012ag}.
In 2015, the CMS Collaboration at LHC confirmed this ratio with newly measured value as~\cite{CMS:2014oqy}
\beq
R^{\prime\; {\rm Exp}}_{3\pi/\pi}
&\equiv&
\frac{{\cal B}(B_c^+ \to J/\psi \pi^+\pi^-\pi^+)}{{\cal B}(B_c^+ \to J/\psi \pi^+)}
= 2.55 \pm 0.87\;.
\label{eq:ex-r3pp-2015-CMS}
\eeq
In the above two ratios, the statistic and systematic errors have been added in quadrature.
By combining these two measurements, the Heavy Flavor Averaging Group gave the averaged ratio
as $2.45 \pm 0.40$~\cite{HFLAV:2022pwe}. Theoretically, the decay $B_c^+ \to J/\psi \pi^+ \pi^-
\pi^+$ has been investigated via definition $R_{3\pi/\pi}$ in Refs.~\cite{Likhoded:2009ib,Rakitin:2009ya,Wang:2012vna,Luchinsky:2012rk,Qiao:2012hp}
with different values, however, which are basically consistent with the current data, except for
a smaller value  $1.5$ estimated in~\cite{Rakitin:2009ya}.

Moreover, in 2013, the decay $B_c^+ \to J/\psi K^+ K^- \pi^+$ was also detected for the first time
by the LHCb Collaboration~\cite{LHCb:2013rud}, and the ratio between ${\cal B}(B_c^+ \to J/\psi K^+ K^-
\pi^+)$ and ${\cal B}(B_c^+ \to J/\psi \pi^+)$ was measured to be
\beq
R_{2K\pi/\pi}^{\rm Exp}
&\equiv& \frac{{\cal B}(B_c^+ \to J/\psi K^+ K^- \pi^+)}{{\cal B}(B_c^+ \to J/\psi \pi^+)}
= 0.53 \pm 0.11 \;,
\label{eq:ex-r2kp-2013-LHCb}
\eeq
where various errors have been added in quadrature too. This measurement agrees well with the
available theoretical predictions $0.49$ and $0.47$~\cite{Luchinsky:2013yla}
corresponding to the resonance approximation with contribution of $a_1$.

Very recently, the LHCb Collaboration studied the $B_c$-meson decaying to charmonia plus multihadron final
states and reported the ratio between the branching fractions of $B_c^+ \to J/\psi K^+ K^- \pi^+$ and $B_c^+ \to J/\psi \pi^+ \pi^- \pi^+$, largely proceeding via
$a_1^+ \to K^+ \bar K^{*0} \to K^+ K^- \pi^+$ and $a_1^+ \to \rho^0 \pi^+ \to \pi^+ \pi^- \pi^+$, respectively,
as follows~\cite{LHCb:2021tdf},
\beq
R_{2K\pi/3\pi}^{\rm Exp} &\equiv&
\frac{{\cal B}(B_c^+ \to J/\psi K^+ K^- \pi^+)}
{{\cal B}(B_c^+ \to J/\psi \pi^+ \pi^- \pi^+)}
= 0.185 \pm 0.014\;,
\label{eq:ex-r2kp3p-2021-lhcb}
\eeq
with the statistic and systematic uncertainties being added in quadrature. This ratio deviates slightly from $0.22 \pm 0.06$ deduced by the previous LHCb measurements as presented in Eqs.~(\ref{eq:ex-r2kp-2013-LHCb}) and (\ref{eq:ex-r3pp-2012-LHCb}).

    \item
In 2013, the decay $B_c^+ \to J/\psi K^+$ was observed for the first time by the LHCb Collaboration
\cite{LHCb:2013hwj}. The ratio between the branching fractions of $B_c^+ \to J/\psi K^+$ and $B_c^+
\to J/\psi \pi^+$ was measured to be
\beq
R_{K/\pi}^{\rm Exp} &\equiv&
\frac{{\cal B}(B_c^+ \to J/\psi K^+)}{{\cal B}(B_c^+ \to J/\psi \pi^+)}
= 0.069 \pm 0.020\;,
\label{eq:ex-rkpi-2013-LHCb}
\eeq
and, subsequently, this ratio was updated in 2016 with a good precision as~\cite{LHCb:2016vni},
\beq
R_{K/\pi}^{\prime\; {\rm Exp}} &\equiv&
\frac{{\cal B}(B_c^+ \to J/\psi K^+)}{{\cal B}(B_c^+ \to J/\psi \pi^+)}
=0.079 \pm 0.008\;.
\label{eq:ex-rkpi-2016-LHCb}
\eeq
The ratio $R_{K/\pi}^{\prime\; {\rm Exp}}$ supersedes the previous $R_{K/\pi}^{\rm Exp}$
however still agrees well with each other.
Notice that the errors from different sources in the above two ratios have been added in quadrature.
Interestingly, the theoretical predictions for this ratio locate at a broad region of $[0.052, 0.088]$
(See them in Table~\ref{tab:Rs-Refs}),
which implies different understanding on the dynamics in different formalisms for these two $B_c$-meson
decays. That is to say, the adequate and complementary studies are still demanded to further clarify the
involved dynamics, in particular, based on certain frameworks of QCD-inspired factorization.
\end{itemize}
Certainly, along with the successful upgrade of LHC, around $10^{10}$ $B_c$-meson events could be
accessed per year. Thus more varieties of $B_c$-meson decays would be measured by the upcoming
experiments at the ongoing LHC, even the forthcoming Circular Electron-Positron
Collider (CEPC).

Motivated by the above-mentioned observations and near future rich measurements on the $B_c$-meson
decays, we propose to study the decays $B_c^+ \to J/\psi M^+$,
in which, $M$ denotes the pseudoscalar ($P$), vector ($V$), axial-vector ($A$),
scalar ($S$) and tensor ($T$) mesons composed
of light quarks, in a comprehensive manner within
the improved perturbative QCD(iPQCD) formalism at leading order~\cite{Liu:2018kuo,Liu:2020upy}.
It is worth emphasizing that the resummation formula adopted in the
conventional PQCD approach to $B_c$-meson decays~\cite{Sun:2008ew,Rui:2014tpa,Rui:2016opu}
is not appropriate. The conventional PQCD formalism has recently been improved by
taking into account the finite charm quark mass effects through $k_T$ resummation at the
next-to-leading-logarithm accuracy~\cite{Liu:2020upy}. The resultant Sudakov factor $s_c(Q,b)$
makes the framework for the $B_c$-meson and $B$-meson decays into charmonia plus light mesons
really complete at leading order. So far, partial decay modes of $B_c^+ \to J/\psi M^+$ such as
$B_c^+ \to J/\psi P^+$ and $B_c^+ \to J/\psi V^+$, even $B_c^+ \to J/\psi a_1^+$, have been
investigated in many different models or methods based on factorization assumptions and
the relative ratios of their branching fractions between the different $B_c^+$ decays
are collected in Table~\ref{tab:Rs-Refs}.
%%================================================
\begin{table}[hbt]
\caption{ Relative ratios from the branching fractions for the decays $B_c^+ \to J/\psi P^+$, $B_c^+ \to J/\psi V^+$, and $B_c^+ \to J/\psi a_1^+$ in the literature at both aspects of theory and experiment.}
\label{tab:Rs-Refs}
\begin{center}\vspace{-0.3cm}{\footnotesize
\begin{tabular}[t]{c|c|c|c|c|c|c|c|c|c|c|c|c|c|c|c|c}
\hline\hline
Ratios &\cite{Chang:1992pt}& \cite{Liu:1997hr}&\cite{Colangelo:1999zn}& \cite{AbdElHady:1999xh} &\cite{Verma:2001hb}  &\cite{Ebert:2003cn} & \cite{Ivanov:2006ni} & \cite{Hernandez:2006gt}& \cite{Naimuddin:2012dy} &\cite{Kar:2013fna} &\cite{Qiao:2012hp}  &\cite{Ke:2013yka}  & \cite{Rui:2014tpa} &\cite{Issadykov:2018myx}& \cite{Cheng:2021svx}
& Data
 \\
\hline
$R_{K/\pi}$ & 0.077& 0.076& 0.052  & 0.074  & 0.049 & 0.082& 0.076 & 0.079 & 0.088 & $\cdots$ & 0.075 &0.079 &0.082 & $0.076^{+0.015}_{-0.015}$   & $0.075^{+0.005}_{-0.005}$
& $0.079^{+0.008}_{-0.008}$
\\
\hline
$R_{K^*/\rho}$ &0.054& 0.054 & 0.054 & 0.057 &0.038 & 0.063 & 0.057 & 0.058 & 0.050 & $\cdots$ & 0.053  & $\cdots$
& 0.057  &0.059& 0.056
& $\cdots$
\\
\hline
$R_{\rho/\pi}$ &3.01 & 3.22& 2.85 & 2.85  &19.31 & 2.62& 2.88 & 3.16  & $\cdots$  &5.29 & 2.77& $\cdots$
& 3.31  &3.52& 5.65
& $\cdots$
\\
\hline
$R_{K^*/\pi}$ &0.16 & 0.17& 0.15 & 0.16  & 0.73 & 0.16& 0.16 & 0.18 & $\cdots$ &0.26 & 0.15 &0.16
& 0.19  &0.21& 0.32
& $\cdots$
\\
\hline
$R_{a_1/\pi}$ & $\cdots$ & $\cdots$ & 4.0 & $\cdots$ & $\cdots$ & $\cdots$ & $\cdots$ & $\cdots$ & $\cdots$ & $\cdots$ &5.5& $\cdots$ & $\cdots$ & $\cdots$ & $\cdots$
& $\cdots$
\\
\hline \hline
\end{tabular}}
\end{center}
\end{table}
%%=========================================================
But, it is indicated that different branching
fractions with large discrepancy for these $B_c^+ \to J/\psi P^+$, $B_c^+ \to J/\psi V^+$,
and $B_c^+ \to J/\psi a_1^+$ channels appear, though the ratios among these branching
fractions are comparable to each other, even to data. As discussed in~\cite{Xiao:2013lia},
the $B_c^+ \to J/\psi P^+$ decays, as well as the $B_c^+ \to J/\psi V^+$ ones, are predominated by
the factorizable emission amplitudes, in association with the negligible
nonfactorizable emission ones.
It means that the theoretical predictions for $R_{K/\pi}$ in various
kinds of models and methods should be consistent with each other, as naively anticipated in
factorization ansatz. That is to say, if the $B_c^+ \to J/\psi V^+$ decays
are basically governed by the longitudinal polarization contributions, then $R_{K^*/\rho}$ is
expected to be close to $R_{K/\pi}$. Certainly,  they cannot tell us more dynamics in the
related decays, even if $R_{K/\pi}^{\rm Theo}$ and $R_{K^*/\rho}^{\rm Theo}$
agree well with those at experiments.
The fact is that the decay amplitude in the decays like
$B_c^+ \to J/\psi P^+$, $B_c^+ \to J/\psi V^+$, even $B_c^+ \to J/\psi a_1^+$
can be approximately written into the product of decay constant and transition form factor as
described in naive factorization, then the above-mentioned ratios can be further
written as the ratio of squared decay constants
multiply by the ratio of squared Cabibbo-Kaboyashi-Maskawa (CKM) matrix elements. Then,
for example, the relation of the branching factions between the $B_c^+ \to J/\psi K^+$
and $B_c^+ \to J/\psi \pi^+$ decays could be naively derived as,
\beq
R_{K/\pi} &\equiv& \frac{{\cal B}(B_c^+ \to J/\psi K^+)}{{\cal B}(B_c^+ \to J/\psi \pi^+)}
\simeq
\frac{|V_{us}|^2}{|V_{ud}|^2}\cdot
\frac{f_K^2}{f_\pi^2}
\sim 0.081
\;,
\label{eq:naiver}
\eeq
with $|V_{us}|= 0.2265, |V_{ud}| = 0.9740, f_K = 0.16$~GeV, and $f_\pi = 0.131$~GeV
\cite{ParticleDataGroup:2020ssz}. This naive expectation
agrees perfectly with the latest measurements
as shown in Eq.~(\ref{eq:ex-rkpi-2016-LHCb}) indeed.
Notice that, however, for the decays with suppressed or vanished factorizable-emission amplitudes
while with enhanced nonfactorizable emission ones, e.g., $B_c^+ \to J/\psi S^+$,
$B_c^+ \to J/\psi T^+$, etc., one should go beyond naive factorization to explore the rich but
complicated dynamics within the factorization framework based on QCD. We can then understand deeply
the perturbative and nonperturbative QCD dynamics involved in these $B_c$-meson decays.

%%%
%%%%%%%%%%%%%%%%% II. Formalism %%%%%%%%%%%%%%%%%%%%%%%%%%%%%%%%
%%%

%\section{ %Formalism and
%perturbative calculations of \boldmath{$B_c^+ \to J/\psi L^+$}}\label{sec:form}

The $B_c$ meson is treated as a heavy-light system~\cite{Liu:2018kuo} and the related decays are analyzed
in its rest frame with momentum $P_{1}= m_{B_c}/\sqrt{2}(1, 1, {\bf 0}_{T})$ in the
light-cone coordinates. Then, for $B_c^+ \to J/\psi M^+$ decays, $M$ and $J/\psi$ mesons
are assumed to move correspondingly in the plus and minus $z$-directions carrying the
momenta $P_2$ and $P_3$ as,
\beq
P_{2}&=& \frac{m_{B_c}}{\sqrt{2}}(1-r_{3}^{2}, r_{2}^{2}, {\bf 0}_{T})\;,
\qquad
P_{3}=\frac{m_{B_c}}{\sqrt{2}}(r_{3}^{2}, 1-r_{2}^{2}, {\bf 0}_{T})\;,
\label{eq:momenta}
\eeq
associated with polarization vectors $\epsilon_2$ and $\epsilon_3$ in the longitudinal
($L$) and transverse ($T$) polarizations, if $M$ is $V$ or $A$, as,
\beq
\epsilon_{2}(L)&=& \frac{1}{\sqrt{2(1-r_{3}^{2})} r_{2}}(1-r_{3}^{2},
-r_{2}^{2}, {\bf 0}_{T}),
\qquad
\epsilon_{2}(T)=(0,0, {\bf 1}_{T})  \;,
\label{eq:polvec-psi}
\\
\epsilon_{3}(L)&=&\frac{1}{\sqrt{2(1-r_{2}^{2})} r_{3}}(-r_{3}^{2}, 1-r_{2}^{2}, {\bf 0}_{T}),
\qquad
\epsilon_{3}(T)=(0,0, {\bf 1}_{T})\;,
\label{eq:polvec-L}
\eeq
where the ratios $r_{2}=m_{M}/m_{B_c}$ and $r_{3}=m_{J/\psi}/m_{B_c}$, and those two polarization
vectors (The capital $L$ and $T$ in the parentheses describe the longitudinal and transverse
polarizations, respectively. Not to be confused with the abbreviation $T$ for tensors. )
satisfy $P \cdot \epsilon = 0$ and $\epsilon^2 =-1$~\footnote{ As described in Ref.
\cite{Liu:2017cwl}, since only three helicities $\ell = 0, \pm 1$ contribute to the $B_c^+
\to J/\psi T^+$ modes, the involved light tensor meson can then be treated
as a vector-like meson with tensor meson mass. In other words, the polarization tensor of tensor meson
can be constructed through the spin-1 polarization vector of vector meson~\cite{Berger:2000wt}.
A new polarization vector $\epsilon_T$ for tensor meson can then be read as
$\epsilon_T(L) = \sqrt{\frac{2}{3}} \epsilon(L)$ and
$\epsilon_T(T)= \sqrt{\frac{1}{2}} \epsilon(T)$~\cite{Datta:2007yk}.~\label{fnt:footnote1}}.
Notice that, due to conservation of the angular momentum, only the longitudinal polarization
vector $\epsilon_{3L}$ of $J/\psi$ is required in the decays $B_c^+ \to J/\psi P^+$ and
$B_c^+ \to J/\psi S^+$. We stress that, due to small contributions
around $5\%$ to the $B_c^+ \to J/\psi M^+$ branching fractions, the terms proportional
to $r_2^2$ and $r_2^4$ will be safely neglected in the numerators of factorization formulas.
The momenta of the spectator quarks in the involved hadrons are parametrized as
\beq
k_1 &=& (x_1 P_1^+, x_1 P_1^-, {\bf k}_{1T}) \;,
\qquad
k_2 = (x_2 P_2^+, x_2 P_2^-, {\bf k}_{2T}) \;,
\qquad
k_3 = (x_3 P_3^+, x_3 P_3^-, {\bf k}_{3T}) \;,
\eeq
where $x_i(i=1,2,3)$ is the momentum fraction of valence quark in the involved mesons.

The $B_c^+ \to J/\psi M^+$ decay amplitude in the iPQCD formalism
can therefore be conceptually written as follows,
\beq
A(B_c^+ \to J/\psi M^+) &\sim &\int\!\! d x_1 d
x_2 d x_3 b_1 d b_1 b_2 d b_2 b_3 d b_3
\non && \cdot {\mathrm{Tr}}
\left [ C(t) \Phi_{B_c}(x_1, b_1) \Phi_{M}(x_2, b_2)
\Phi_{J/\psi}(x_3, b_3) H(x_i, b_i, t) e^{-S(t)} \right ]\;,
\label{eq:a2}
\eeq
where $b_i$ is the conjugate space coordinate of transverse momentum $k_{iT}$;
$t$ is the largest running energy
scale in hard kernel $H(x_i,b_i,t)$; Tr denotes the trace over Dirac and SU(3) color indices;
$C(t)$ stands for the Wilson coefficients including the large logarithms $\ln (m_W/t)$~\cite{Keum:2000ph};
and $\Phi$ is the wave function describing the hadronization of quarks and antiquarks to the meson.
The Sudakov factor $e^{-S(t)}$ arises from $k_T$ resummation, which provides a strong suppression
on the long distance contributions in the small $k_T$(or large $b$) region~\cite{Botts:1989kf}
The detailed discussions for $e^{-S(t)}$ can be easily found in the original Refs.
\cite{Botts:1989kf,Liu:2018kuo,Liu:2020upy}.
Thus, with Eq.~(\ref{eq:a2}), we can give the convoluted amplitudes of the decays $B_c^+ \to J/\psi M^+$
explicitly through the evaluations of the hard kernel $H(x_i,b_i,t)$ at leading order in the $\alpha_s$
expansion with the iPQCD formalism.

%\subsection{Hadron wave functions}
%\label{ssec:wf}

The wave function for $B_c$ meson with a heavy-light structure
can generally be defined as~\cite{Liu:2018kuo,Keum:2000ph,Lu:2002ny}
\beq
\Phi_{B_c}(x, k_T) &=& \frac{i }{\sqrt{2N_c}}
\biggl\{(\psl +m_{B_c})\gamma_5
 \phi_{B_c}(x, k_T) \biggr\}_{\alpha\beta}\;,
\label{eq:def-bq}
\eeq
where $\alpha,\beta$ are the color indices; $P$ is the momentum of $B_c$ meson;
$N_c =3$ is the color factor; and $x$ and $k_T$ are the momentum fraction and
intrinsic transverse momentum of charm quark in the $B_c$ meson;
$\phi_{B_c}(x,k_T)$ is the $B_c$-meson leading-twist distribution amplitude.

For the vector $J/\psi$ meson, its wave function has been studied within the nonrelativistic QCD
approach~\cite{Bondar:2004sv}. The longitudinal and transverse wave functions have been derived as,
\beq
\Phi_{J/\psi}^{L}(x) &=& \frac{1}{\sqrt{2N_{c}}}
\biggl\{m_{J/\psi}\epsl_{L}\phi_{J/\psi}^{L}(x)
+\epsl_{L}\psl\ \phi_{J/\psi}^{t}(x)\biggl\}_{\alpha\beta}\;,
\label{eq:wf-psi-L}\\
\Phi_{J/\psi}^{T}(x) &=& \frac{1}{\sqrt{2N_{c}}}
\biggl\{m_{J/\psi}\epsl_{T}\phi_{J/\psi}^{v}(x)
+\epsl_{T}\psl \ \phi_{J/\psi}^{T}(x)\biggl\}_{\alpha\beta}\;.
\label{eq:wf-psi-T}
\eeq
Here, $x$ describes the distribution of charm quark momentum in $J/\psi$ meson,
$\epsilon_{L}$ and $\epsilon_{T}$ are the two polarization vectors of $J/\psi$,
and $\phi^{L}_{J/\psi}(x)$
and $\phi^{T}_{J/\psi}(x)$ are the twist-2 distribution amplitudes, while $\phi^{t}_{J/\psi}(x)$ and
$\phi^{v}_{J/\psi}(x)$ are the twist-3 ones.

The light-cone wave functions including distribution amplitudes for light pseudoscalars, scalars, vectors,
axial-vectors, and tensors have been given in the QCD sum rules up to twist-3. They are collected as follows:
\begin{itemize}
    \item {For $P$ and $S$ mesons~\cite{Chernyak:1983ej,Ball:1998tj,Braun:2004vf,Cheng:2005nb,Li:2008tk} },
\beq
\Phi_P(x) &=& \frac{i}{\sqrt{2 N_c}} \gamma_5
\biggl\{\psl \phi_P^A(x)+ m_0^P \phi_P^P(x) + m_0^P (\nsl \vsl -1)
\phi_P^T(x)\biggr\}_{\alpha\beta}\;,
\eeq
 and
\beq
\Phi_S(x) &=& \frac{i}{\sqrt{2 N_c}}
\biggl\{\psl \phi_S(x)+ m_S \phi_S^S(x) + m_S (\nsl \vsl -1)
\phi_S^T(x)\biggr\}_{\alpha\beta}\;,
\eeq
with $m_S (m_0^P)$ denoting (chiral) mass of light (pseudoscalars) scalars, $n=(1, 0, {\bf 0}_T)$
and $v=(0,1,{\bf 0}_T)$ being the light-like dimensionless vectors on the light-cone.
And $\phi_P^A(x)$ and $\phi_S(x)$ are the leading-twist distribution amplitudes,
while $\phi_P^{P, T}(x)$ and $\phi_S^{S, T}(x)$ are the twist-3 ones.

    \item {For $V$ and $A$ mesons with
    polarizations~\cite{Ball:vector,Ball:2007rt,Yang:2007zt,Li:2009tx}, }
\beq
\Phi^{L}_{V}(x) &=&  \frac{1}{\sqrt{2 N_c}}  \biggl\{ m_{V}\, {\epsl}_V^L \,\phi_{V}(x)
+ {\epsl}_V^L \, \psl\,\phi_{V}^t(x)  + m_{V}\, \phi_{V}^s(x) \biggr\}_{\alpha\beta}\;,
 \\
\Phi^{T}_{V}(x) &=&  \frac{1}{\sqrt{2 N_c}}  \biggl\{ m_{V}\, {\epsl}_V^T\, \phi_{V}^v(x)
+ {\epsl}_V^T\, \psl\, \phi_{V}^T(x)+m_{V}
i\epsilon_{\mu\nu\rho\sigma}\gamma_5\gamma^\mu {\epsl}_V^{T\nu}
n^\rho v^\sigma \phi_{V}^a(x) \biggr\}_{\alpha\beta}\;,
 \eeq
 and
 \beq
 \Phi_A^L (x) &=& \gamma_5 \Phi_V^L (x)\;, \qquad
 \Phi_A^T (x) =  \gamma_5 \Phi_V^T (x)\;,
 \eeq
where $m, P$ and $\epsilon$ are the mass, the momentum and the polarization vector for
(axial-)vector mesons, $\phi(x)$ and $\phi^{T}(x)$ are the leading-twist distribution amplitudes,
and $\phi^{t,s}(x)$ and $\phi^{v,a}(x)$ are the twist-3 ones,
and $x$ is the momentum fraction of quark carrying in the
(axial-)vector mesons.

   \item {For $T$ mesons with polarizations~\cite{Cheng:2010hn,Wang:2010ni}, }
\beq
\Phi^{L}_{T}(x) &=&  \frac{1}{\sqrt{2 N_c}}
   \biggl\{ m_{T}\, {\epsl}_T^L \,\phi_{T}(x)  +
 {\epsl}_T^L \, \psl\,\phi_{T}^t(x)  + m_{T}\, \phi_{T}^s(x) \biggr\}_{\alpha\beta}\;,
 \\
% \eeq
% %%
% \beq
\Phi^{T}_{T}(x)
 &=&  \frac{1}{\sqrt{2 N_c}}
  \biggl\{ m_{T}\, {\epsl}_T^T\, \phi_{T}^v(x) +
{\epsl}_T^T\, \psl\, \phi_{T}^T(x)+m_{T}
i\epsilon_{\mu\nu\rho\sigma}\gamma_5\gamma^\mu {\epsl}_T^{T\nu}
n^\rho v^\sigma \phi_{T}^a(x) \biggr\}_{\alpha\beta}\;,
 \eeq
with the tensor meson mass $m_T$, the twist-2 distribution amplitudes $\phi_T(x)$ and $\phi_T^T(x)$,
the twist-3 distribution amplitudes $\phi_T^{t,s}(x)$ and $\phi_T^{v,a}(x)$, the momentum $P$ and
polarization vector $\epsilon_T$ satisfying $P\cdot \epsilon_T =0$, and the momentum fraction
$x$ carried by quark in the tensor meson.
\end{itemize}
In the above wave functions with polarizations, we adopt the convention $\epsilon^{0123}=1$ for the
Levi-Civit$\grave{a}$ tensor $\epsilon^{\mu\nu\alpha\beta}$. Notice that the explicit expressions for
all the above-mentioned distribution amplitudes with their masses, decay constants and Gegenbauer moments
can be found later in Appendix~\ref{app:Dis-Amps}.

%\subsection{Perturbative calculations} \label{ssec:pcalc}

For the $B_c^+ \to J/\psi M^+$ decays induced by the $\bar b \to \bar c$ transition
at the quark level, the related weak effective Hamiltonian $H_{{\rm eff}}$ can be written
as~\cite{Buchalla:1995vs}
\beq
H_{\rm eff}\, &=&\, {G_F\over\sqrt{2}}
\biggl\{ V^*_{cb}V_{uq} [ C_1(\mu) O_1(\mu)
+C_2(\mu) O_2(\mu) ] \biggr\}+ {\rm H.c.}\;,
\label{eq:heff}
\eeq
with the Fermi constant $G_F=1.16639\times 10^{-5}{\rm GeV}^{-2}$, the CKM matrix elements $V$,
and the Wilson coefficients $C_i(\mu)$ at the renormalization scale $\mu$. The local four-quark
tree operators $O_1$ and $O_2$ are written as
\beq
O_1 \, &=&\,
\bar{q}_\alpha \gamma_\mu (1 - \gamma_5) u_\beta\; \bar{c}_\beta \gamma_\mu (1 - \gamma_5) b_\alpha \;,
\qquad
O_2 \, =\, \bar{q}_\alpha \gamma_\mu (1 - \gamma_5) u_\alpha\; \bar{c}_\beta \gamma_\mu (1 - \gamma_5) b_\beta \;,
\label{eq:operators}
\eeq
where $q$ denotes the light down quark $d (s)$ for the $\Delta S =0\ (1)$,
namely, CKM-favored (suppressed) processes with
$S$ being the strange number (Not to be confused with the abbreviation $S$ for scalars).

%%%%=============================================================
\begin{figure}[!!htb]
\centering
\begin{tabular}{l}
\includegraphics[width=0.95\textwidth]{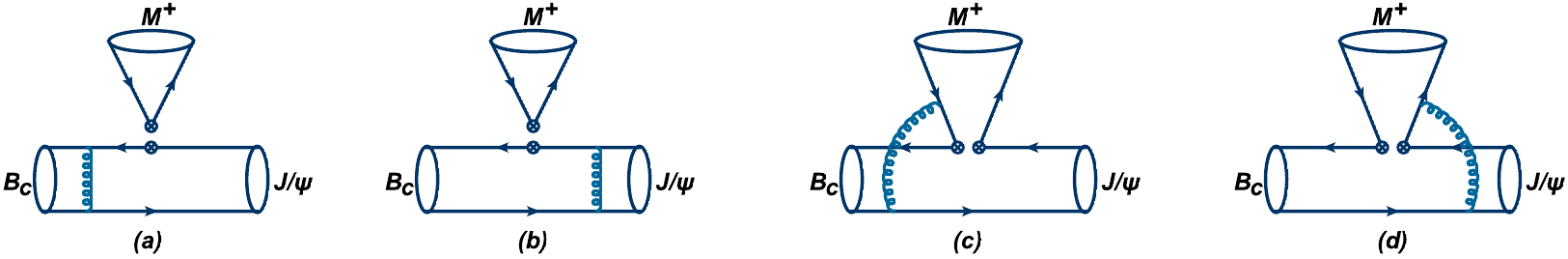}
\end{tabular}
\caption{(Color online) Leading order Feynman diagrams for the decays $B_c^+ \to J/\psi M^+$
in the iPQCD formalism. }
\label{fig:fig1}
\end{figure}
%%%%==============================================================

The related Feynman diagrams for the decays $B_c^+ \to J/\psi M^+$ in the iPQCD formalism at
leading order are illustrated in Fig.~\ref{fig:fig1}. As presented
[see Eqs.~(31)-(46) for details] in Ref.~\cite{Liu:2018kuo}, we have given the factorization
formulas and analytic $B_c^+ \to J/\psi \pi^+$ decay
amplitudes with all elements. The similar calculations could be repeated for the rest
$B_c^+ \to J/\psi M^+$ decay modes in this work. Hereafter, for the sake of simplicity, we will use
$F_{e}$ and $M_{e}$ to describe the factorizable emission and the nonfactorizable emission
amplitudes induced by the $(V-A)(V-A)$ operators in these types of $B_c^+ \to J/\psi M^+$
decays. Furthermore, in light of the successful clarification of most branching ratios and
polarization fractions in the $B \to VV$ decays by keeping the terms proportional
to $r^2_V=m_V^2/m_B^2$ in the denominator of propagators for virtual quarks and gluons with
the PQCD approach~\cite{Zou:2015iwa}, we will follow this treatment in the present work.
That is, we will retain the terms like $r_2^2$ and $r_3^2$ in dealing with the denominators
of factorization formulas for the decays $B_c^+ \to J/\psi M^+$, which could be
examined by future measurements to further clarify its universality. The related factorization
formulas can be found in Appendix~\ref{app:facformulas}.

The $B_c^+ \to J/\psi M^+$ decay amplitude can thus be decomposed into
\beq
A^{(\sigma)}(B_c^+ \to J/\psi M^+) &=& V_{cb}^* V_{uq} ( F_e^{(\sigma)} \cdot f_{M} +  M_e^{(\sigma)} )\;,
\label{eq:DecAmp}
\eeq
with $\sigma = L$ for the modes $B_c^+ \to J/\psi P^+$ and $J/\psi S^+$ involving contributions
from only longitudinal polarization while $\sigma = L, N, T$ for the channels $B_c^+ \to J/\psi V^+$,
$J/\psi A^+$, and $J/\psi T^+$ containing contributions from longitudinal, normal, and transverse
polarizations, which result in the formulas for calculating branching fractions of the decays $B_c^+ \to
J/\psi M^+$ as follows,
\begin{itemize}
\item { for the decays $B_c^+ \to J/\psi M^+$ with $M = P$ and $S$, }
\beq
{\cal B}(B_c^+ \to J/\psi M^+)&\equiv&
\tau_{B_c}\cdot \Gamma(B_c^+ \to J/\psi M^+)
\non
&=& \tau_{B_c}\cdot\frac{G_{F}^{2} m^{3}_{B_c}}
{32 \pi}\cdot \Phi(r_2, r_3)
\cdot |A(B_c^+ \to J/\psi M^+) |^2\;,
\label{eq:br-def}
\eeq
where $\tau_{B_c}$ is the lifetime of $B_c$ meson and $\Phi(r_2, r_3)$
is the phase space
factor of $B_c^+ \to J/\psi M^+$ decays with $\Phi(x, y) \equiv
\{ [1-(x+y)^2] \cdot [1-(x-y)^2] \}^{1/2}$~\cite{Fleischer:2011au}.

\item {for the decays $B_c^+ \to J/\psi M^+$ with $M = V$, $A$, and $T$, }
\beq
{\cal B}(B_c^+ \to J/\psi M^+)&\equiv&
\tau_{B_c}\cdot \Gamma(B_c^+ \to J/\psi M^+)
\non
&=& \tau_{B_c}\cdot\frac{G_{F}^{2}|\bf{P_c}|}{16 \pi m^{2}_{B_c} }
\sum_{\sigma=L,N,T} A^{\sigma\; \dagger }(B_c^+ \to J/\psi M^+)
A^{\sigma}(B_c^+ \to J/\psi M^+)\;,
\label{dr1}
\eeq
where $|\bf{P_c}|\equiv |\bf{P_{2z}}|=|\bf{P_{3z}}|$ is the momentum of either the outgoing $M$
meson or $J/\psi$ meson and $A^{\sigma}(B_c^+ \to J/\psi M^+)$ denotes the decay amplitudes with
helicities for the decays $B_c^+ \to J/\psi V^+$, $J/\psi A^+$, and $J/\psi T^+$, respectively.
\end{itemize}

%\section{Numerical Results and Discussions} \label{sec:randd}

Now we turn to carry out the numerical calculations and make phenomenological analyses. In
numerical calculations, central values of the input parameters will be used implicitly
unless otherwise stated. The relevant QCD scale~({\rm GeV}), masses~({\rm GeV}), and
$B_c$-meson lifetime ({\rm ps}) are the following
\cite{Keum:2000ph,ParticleDataGroup:2020ssz}
\beq
 \Lambda_{\overline{\mathrm{MS}}}^{(f=4)} &=& 0.250\; ,
 \qquad m_W = 80.41\;,
 \qquad m_{B_c} = 6.275 \;,
 \qquad m_{J/\psi}  = 3.097\;,
 \non
   \tau_{B_c} &=& 0.507\;,
 \qquad  m_b = 4.8 \;,
 \qquad  m_{c}= 1.5 \;.
\label{eq:mass}
\eeq

For the CKM matrix elements, the Wolfenstein parametrization up to ${\cal O}(\lambda^4)$ is adopted
\cite{Wolfenstein:1983yz},
\beq
V_{\mbox{{\scriptsize CKM}}} &=& \left(\begin{array}{ccc}
1-\frac{1}{2}\lambda^2 \qquad & \lambda \qquad & A\lambda^3(\rho-i\eta) \\
-\lambda \qquad & 1-\frac{1}{2}\lambda^2 \qquad & A\lambda^2\\
A\lambda^3(1-\rho-i\eta) \qquad & -A\lambda^2 \qquad & 1
\end{array}\right)+{\cal O}(\lambda^4)\;,
\label{eq:wolfenstein}
\eeq
with the updated parameters $A=0.790$ and $\lambda=0.2265$~\cite{ParticleDataGroup:2020ssz}.

\subsection{{\boldmath  $ B_c^+ \to J/\psi (P^+, V^+)$}}
\label{ssec:psipv}

As inferred from the literature, the decays $B_c^+ \to J/\psi P^+$ and $J/\psi V^+$ have been
studied extensively in many different approaches and methods, though with different individual
branching fractions, for example, see Refs.~\cite{Chang:1992pt,Liu:1997hr,Colangelo:1999zn,AbdElHady:1999xh,Verma:2001hb,
Ebert:2003cn,Ivanov:2006ni,Hernandez:2006gt,Naimuddin:2012dy,Kar:2013fna,Qiao:2012hp,Ke:2013yka,Rui:2014tpa,Issadykov:2018myx,
Cheng:2021svx}. The {\it CP}-averaged branching fractions of the $B_c^+ \to J/\psi P^+$ decays
in the iPQCD formalism are
\beq
{\cal B}(B_c^+ \to J/\psi \pi^+)&=&
1.17
%^{+0.32}_{-0.29}
^{+0.31}_{-0.23}(\beta_{B_c})
^{+0.08}_{-0.08}(f_M)
%^{+0.02}_{-0.02}(f_{B_c})
%^{+0.08}_{-0.08}(f_{J/\psi})
^{+0.00}_{-0.00}(a_\pi)
%^{+0.01}_{-0.01}(a_t)
\times 10^{-3} \;,
\\
{\cal B}(B_c^+ \to J/\psi K^+) &=&
8.68
%^{+0.25}_{-0.19}
^{+2.32}_{-1.73}(\beta_{B_c})
^{+0.64}_{-0.62}(f_M)
%^{+0.18}_{-0.18}(f_{B_c})
%^{+0.61}_{-0.59}(f_{J/\psi})
^{+0.66}_{-0.63}(a_K)
%%^{+0.66+0.00}_{-0.63-0.02}(a_K)
%^{+0.08}_{-0.12}(a_t)
\times 10^{-5}\;,
\eeq
where the dominant error arises from the shape parameter $\beta_{B_c}$ in $B_c$-meson
distribution amplitude. In the $B_c^+ \to J/\psi K^+$ mode, the uncertainties from the
combined decay constants of $B_c$ and $J/\psi$ and from the SU(3)-flavor symmetry breaking
factor $a_1^K$ can compete with each other. These two branching
ratios are around ${\cal O}(10^{-3})$ and ${\cal O}(10^{-4})$, respectively, within uncertainties.
Although the individual $B_c^+ \to J/\psi \pi^+$ branching fraction is not definitely available yet,
our iPQCD prediction for its value agrees generally with most of the predictions from various
models and methods already presented in the literature, in particular, with that given in
QCD factorization approach~\cite{Sun:2007ei}, and presented very recently in covariant confined
quark model~\cite{Issadykov:2018myx} and light-cone sum rules approach~\cite{Cheng:2021svx}.
Furthermore, our predictions about the $B_c^+ \to J/\psi P^+$ branching fractions agree well with
those presented in Refs.~\cite{Issadykov:2018myx,Cheng:2021svx}.

The ratio between the {\it CP}-averaged branching fractions of $B_c^+ \to J/\psi K^+$ and
$B_c^+ \to J/\psi \pi^+$ in the iPQCD formalism is therefore given theoretically as,
\beq
R^{\rm Theo}_{K/\pi}&\equiv& \frac{{\cal B}(B_c^+ \to J/\psi K^+)}
{{\cal B}(B_c^+ \to J/\psi \pi^+)}=
%7.42
%^{+0.01+0.03+0.01+0.56+0.00}_{-0.03-0.03-0.00-0.54-0.04}
%^{+0.56}_{-0.54}
%\times 10^{-2}
0.074^{+0.006}_{-0.005}\;,
\label{eq:rkpi-th}
\eeq
which agrees well with the very recent predictions in Refs.~\cite{Issadykov:2018myx,Cheng:2021svx} and the
latest measurement as shown in Eq.~(\ref{eq:ex-rkpi-2016-LHCb}) within errors.
This natural agreement is trivial because these two $B_c^+ \to J/\psi P^+$ modes are absolutely
dominated by the factorizable emission diagrams while with dramatically small nonfactorizable emission
contributions due to the considerable cancelation under the isospin limit.
It is clear to see that $R^{\rm Theo}_{K/\pi}$
predicted in iPQCD formalism is consistent with the naive expectation  $R_{K/\pi}$ as described
in Eq.~(\ref{eq:naiver}), besides in good
consistency with the measured one. The slight deviation between $R^{\rm Theo}_{K/\pi}$ and
$R_{K/\pi}$ arises from the SU(3)-flavor symmetry breaking effects in the leading-twist
distribution amplitude of kaon, due to the fact that the contributions induced by the
nonfactorizable emission diagrams are proportional to $\phi_P^A(x)$ as shown in Eq.~(33)
of Ref.~\cite{Liu:2018kuo}. Therefore, as a by-product, it could be anticipated that the above
naive expectation is also validated for $B_c$-meson decays into other charmonia, e.g., $\eta_c$,
$\chi_{cJ}(J=0,1,2)$, \ldots, plus a pion or a kaon. However, it is worth stressing that the
measurements on the ratio while without individual decay rate cannot help reveal the
involved dynamics, even further constrain the related hadronic parameters, for example, the
shape parameter $\beta_{B_c}$ in the $B_c$-meson leading-twist distribution amplitude $\phi_{B_c}$.
In other words, the investigations on the related modes with large nonfactorizable
contributions are of great necessity.

Now, we turn to analyze the $B_c^+ \to J/\psi V^+$ decays. The $B_c^+ \to J/\psi V^+$ branching
fractions in the iPQCD formalism can be read as follows:
\beq
{\cal B}(B_c^+ \to J/\psi \rho^+)&=&
3.69
%^{+1.06}_{-0.82}
^{+1.02}_{-0.76}(\beta_{B_c})
^{+0.28}_{-0.27}(f_M)
%^{+0.08}_{-0.07}(f_{B_c})
%^{+0.26}_{-0.25}(f_{J/\psi})
%^{+0.07}_{-0.07}(f_\rho)
%^{+0.00}_{-0.00}(f_\rho^T)
^{+0.01}_{-0.00}(a_\rho)
%%^{+0.01}_{-0.00}(a_2^\parallel)
%%^{+0.00}_{-0.00}(a_2^\perp)
%^{+0.09}_{-0.12}(a_t)
\times 10^{-3}  \;,
\\
{\cal B}(B_c^+ \to J/\psi K^{*+})&=&
2.23
%^{+0.65}_{-0.50}
^{+0.62}_{-0.46}(\beta_{B_c})
^{+0.20}_{-0.18}(f_M)
%^{+0.05}_{-0.04}(f_{B_c})
%^{+0.16}_{-0.15}(f_{J/\psi})
%^{+0.11}_{-0.10}(f_{K^*})
%^{+0.00}_{-0.00}(f_{K^*}^T)
^{+0.02}_{-0.01}(a_{K^*})
%%^{+0.02}_{-0.01}(a_1^\parallel)
%%^{+0.00}_{-0.00}(a_1^\perp)
%%^{+0.01}_{-0.00}(a_2^\parallel)
%%^{+0.00}_{-0.00}(a_2^\perp)
%^{+0.06}_{-0.07}(a_t)
\times 10^{-4} \;,
\eeq
where the errors are dominated by the $B_c$-meson shape parameter $\beta_{B_c}$
and the combination of decay constants of the $B_c$-meson and related vectors.
These predictions are well consistent with those in~Ref.~\cite{Issadykov:2018myx}
within theoretical errors.

Based on the helicity amplitudes, we can define the transversity ones as follows:
\beq
{\cal A}_{L}&=& \xi m^{2}_{B} A^{L}, \qquad
{\cal A}_{\parallel}=\xi \sqrt{2}m^{2}_{B} A^{N}, \qquad
{\cal A}_{\perp}=\xi m_{V} m_{J/\psi}
\sqrt{2(r^{2}-1)} A^{T}\;,
\label{eq:ase}
\eeq
for the longitudinal, parallel, and perpendicular polarizations, respectively, with the
normalization factor $\xi=\sqrt{G^2_{F}|{\bf P_c}|/(16\pi m^2_{B_c}\Gamma)}$ and the ratio
$r=P_{2}\cdot P_{3}/(m_{V}\cdot m_{J/\psi})$. These amplitudes satisfy the relation,
\beq
|{\cal A}_{L}|^2+|{\cal A}_{\parallel}|^2+|{\cal A}_{\perp}|^2 &=& 1 \;.
\eeq
following the summation in Eq.~(\ref{dr1}). Since the transverse-helicity contributions can
manifest themselves through polarization observables, we therefore define {\it CP}-averaged longitudinal
polarization fractions $f_{L}$ as the following,
\beq
f_{L}&\equiv&
\frac{|{\cal A}_{L}|^2}
{|{\cal A}_L|^2+|{\cal A}_{||}|^2+|{\cal
A}_{\perp}|^2}\;
(= |{\cal A}_{L}|^2)\;.
\label{eq:pf}
\eeq
Then the {\it CP}-averaged longitudinal polarization fractions of the $B_c^+ \to J/\psi V^+$ modes can be
presented as follows,
\beq
f_L(B_c^+ \to J/\psi \rho^+)&=&
(89.1^{+0.1}_{-0.1})
%^{+0.1}_{-0.1}(\beta_{B_c})
%%^{+0.0}_{-0.0}(f_M)
%^{+0.0}_{-0.0}(f_{B_c})
%^{+0.0}_{-0.0}(f_{J/\psi})
%^{+0.0}_{-0.0}(f_\rho)
%^{+0.0}_{-0.0}(f_\rho^T)
%^{+0.0}_{-0.0}(a_\rho)
%%^{+0.0}_{-0.0}(a_2^\parallel)
%%^{+0.0}_{-0.0}(a_2^\perp)
%^{+0.0}_{-0.0}(a_t)
\%  \;,
\qquad
f_L(B_c^+ \to J/\psi K^{*+})=
(85.6^{+0.2}_{-0.2})
%^{+0.1}_{-0.1}(\beta_{B_c})
%%^{+0.0}_{-0.0}(f_M)
%^{+0.0}_{-0.0}(f_{B_c})
%^{+0.0}_{-0.0}(f_{J/\psi})
%^{+0.0}_{-0.0}(f_{K^*})
%^{+0.0}_{-0.0}(f_{K^*}^T)
%^{+0.1}_{-0.1}(a_{K^*})
%%^{+0.1}_{-0.1}(a_1^\parallel)
%%^{+0.0}_{-0.0}(a_1^\perp)
%%^{+0.1}_{-0.0}(a_2^\parallel)
%%^{+0.0}_{-0.0}(a_2^\perp)
%^{+0.0}_{-0.1}(a_t)
\% \;.
\eeq
It is found that the decays $B_c^+ \to J/\psi V^+$ are generally governed by the longitudinal
amplitudes in the iPQCD formalism but with slightly different fractions.

The ratio between the {\it CP}-averaged branching fractions of $B_c^+ \to J/\psi K^{*+}$ and
$B_c^+ \to J/\psi \rho^+$ is then obtained as follows,
\beq
R^{\rm Theo}_{K^*/\rho}&\equiv&
\frac{{\cal B}(B_c^+ \to J/\psi K^{*+})}{{\cal B}(B_c^+ \to J/\psi \rho^+)}
= 0.060^{+0.002}_{-0.002}
%6.04
%^{+0.01+0.01+0.01+0.18+0.04+0.02}_{-0.00-0.00-0.00-0.16-0.02-0.00}
%^{+0.19}_{-0.16}  \times 10^{-2}
\;,
\eeq
which roughly meets with the value $R_{K^*/\rho} \sim 0.066$ anticipated by
naive factorization with $f_{K^*} =0.118$~GeV,
$f_\rho = 0.107$~GeV, $|V_{us}| = 0.2265$ and $|V_{ud}| = 0.9740$,
and is indeed close to the ratio $R^{\rm Theo}_{K/\pi}$ presented in
Eq.~(\ref{eq:rkpi-th}) within errors. The ratio $R^{\rm Theo}_{K^*/\rho}$
is expected to be measured at the near future LHC experiments.

With the normalization channel, i.e., $B_c^+ \to J/\psi \pi^+$, therefore the ratios
between the branching fractions of $B_c^+ \to J/\psi V^+$ and $B_c^+ \to J/\psi \pi^+$
predicted for future examination are as follows,
\beq
R^{\rm Theo}_{\rho/\pi}&\equiv&
\frac{{\cal B}(B_c^+ \to J/\psi \rho^+)}{{\cal B}(B_c^+ \to J/\psi \pi^+)}
=
3.15
%^{+0.03+0.02+0.01+0.05+0.01+0.06}_{-0.03-0.00-0.00-0.07-0.00-0.06}
^{+0.09}_{-0.10}
\;,
\qquad
R^{\rm Theo}_{K^*/\pi} \equiv
\frac{{\cal B}(B_c^+ \to J/\psi K^{*+})}{{\cal B}(B_c^+ \to J/\psi \pi^+)}
=
0.19
%^{+0.00+0.00+0.00+0.01}_{-0.00-0.00-0.00-0.01}
^{+0.01}_{-0.01}
\;.
\eeq

\subsection{\boldmath $B_c^+ \to J/\psi A^+$}
\label{ssec:psiav}

The $p$-wave light axial-vectors have been investigated at both experimental and theoretical aspects.
However, the understanding about their internal structure is far from satisfactory~\cite{Du:2022nno}. As presented in
the particle list by Particle Data Group, the nonstrange axial-vectors $a_1$ and $b_1(1235)$ (In the following context,
we will use $b_1$ to denote this state for simplification.) belong to two different types of bound states
in the constituent quark model with quantum numbers $1^3\!P_1(J^{PC} = 1^{++})$ and $1^1\!P_1(J^{PC} =
1^{+-})$, respectively. While for the strange $K_1(1270)$ and $K_1(1400)$ mesons (We will conveniently adopt
$K_1$ and $K_1^\prime$ to denote these two states), these two physical states are considered
as the mixtures of $K_{1A} (1^3\!P_1)$ and $K_{1B} (1^1\!P_1)$ with a single angle $\theta_{K_1}$ due
to the mass difference of the strange and nonstrange light quarks~\cite{Cheng:2011pb},
\beq
\left(
\begin{array}{c} |K_1 \rangle \\ |K_1^\prime \rangle \\ \end{array} \right ) &=&
  \left( \begin{array}{cc}
\sin{\theta_{K_1}} & \hspace{0.28cm} \cos{\theta_{K_1}} \\
 \cos{\theta_{K_1}} & -\sin\theta_{K_1} \end{array} \right )
 \left( \begin{array}{c}  |K_{1A}\rangle\\ |K_{1B} \rangle \\ \end{array} \right )\;,
 \label{eq:mixture}
\eeq
In analogy to the decays $B \to \phi K_1^{(\prime)}$~\cite{Liu:2014dxa}, we take both $\theta_{K_1}
\approx 33^\circ$ and $58^\circ$ into account of the calculations to estimate the branching fractions
because of the currently unknown nature.

Then, for the two $\Delta S =0$ channels, $B_c^+ \to J/\psi a_1^+$ and $B_c^+ \to J/\psi b_1^+$,
the predictions for their branching fractions in the iPQCD formalism can be read as,
\beq
{\cal B}(B_c^+ \to J/\psi a_1^+)&=&
5.90^{+1.63}_{-1.24}(\beta_{B_c})
^{+0.66}_{-0.64}(f_M)
%^{+0.12}_{-0.12}(f_{B_c})
%^{+0.41}_{-0.40}(f_{J/\psi})
%^{+0.50}_{-0.49}(f_{a_1})
^{+0.00}_{-0.00}(a_{a_1})
%^{+0.0}_{-0.0}(a_1^t)
%^{+0.0}_{-0.0}(a_2^p)
%^{+0.14}_{-0.20}(a_t)
\times 10^{-3} \;,
\\
{\cal B}(B_c^+ \to J/\psi b_1^+)&=&
7.93^{+2.43}_{-1.83}(\beta_{B_c})
^{+0.93}_{-0.89}(f_M)
%^{+0.16}_{-0.16}(f_{B_c})
%^{+0.56}_{-0.54}(f_{J/\psi})
%^{+0.72}_{-0.69}(f_{b_1})
^{+3.09}_{-2.59}(a_{b_1})
%^{+0.03}_{-0.03}(b_0^p)
%^{+3.09}_{-2.59}(b_1^p)
%^{+0.00}_{-0.00}(b_2^t)
%^{+0.15}_{-0.12}(a_t)
\times 10^{-4} \;.
\eeq
One can easily find that the $B_c^+ \to J/\psi b_1^+$ branching ratio suffers from large errors
induced by the uncertainties of Gegenbauer moments in the $b_1$-meson distribution amplitudes.
Owing to the nearly vanished decay constant $f_{b_1^+} = f_{b_1} \cdot a_{0b_1}^{\parallel} \sim 0.0005$
(Here, $f_{b_1}$ and $a_{0b_1}^{\parallel}$ are the ``normalization" decay constant and the zeroth
Gegenbauer moment for the meson $b_1^+$, respectively, and they could be found explicitly in
Table IV of Appendix~\ref{app:Dis-Amps}. Note that $a_{0b1}^\parallel =0$ under the isospin limit.),
therefore the factorizable emission
contributions are extremely suppressed, which results in the smaller ${\cal B}(B_c^+ \to J/\psi b_1^+)$
that comes almost from the nonfactorizable emission decay amplitudes. However, the antisymmetric
behavior of the $b_1$-meson leading-twist distribution amplitude changes the destructive interferences
between the two diagrams like Figs.~\ref{fig:fig1}(c) and~\ref{fig:fig1}(d) in the $B_c^+ \to J/\psi a_1^+$
channel into the constructive ones in the $B_c^+ \to J/\psi b_1^+$ mode, which lead to ${\cal B}(B_c^+ \to J/\psi b_1^+)$ around ${\cal O}(10^{-3})$ within large errors.

The ratios between the branching fractions ${\cal B}(B_c^+ \to J/\psi a_1^+/b_1^+)$
and ${\cal B}(B_c^+ \to J/\psi \pi^+)$ can be defined as follows,
\beq
R^{\rm Theo}_{a_1/\pi}&\equiv&
\frac{{\cal B}(B_c^+ \to J/\psi a_1^+)}{{\cal B}(B_c^+ \to J/\psi \pi^+)}=
5.04^{+0.10}_{-0.15}
%^{+0.05+0.02+0.01+0.08}_{-0.08-0.01-0.00-0.13}
%\times 10^{0}
\label{eq:ra1pi}\;,
\qquad
R^{\rm Theo}_{b_1/\pi} \equiv
\frac{{\cal B}(B_c^+ \to J/\psi b_1^+)}{{\cal B}(B_c^+ \to J/\psi \pi^+)}=
0.68^{+0.26}_{-0.22}
%^{+0.02+0.00+0.00+0.26+0.02}_{-0.03-0.00-0.00-0.22-0.01}
%\times 10^{0}
\label{eq:rb1pi}\;,
\eeq
which would help probe these two channels experimentally in the near future.
The large ratio $R_{b_1/\pi}^{\rm Theo}$ needs experimental tests as soon as possible to
provide useful hints to test the reliability of iPQCD formalism
utilized in this type of decays. If the information from experiments
is positive,
then these types of $B_c$-meson decay modes would offer good opportunities to help explore the
shape parameter $\beta_{B_c}$ in the $B_c$-meson distribution amplitude phenomenologically.

As presented in Eqs.~(\ref{eq:ex-r3pp-2012-LHCb}) and~(\ref{eq:ex-r3pp-2015-CMS}), the decay $B_c^+ \to J/\psi a_1^+$
has been studied experimentally through the $B_c^+ \to J/\psi \pi^+ \pi^-
\pi^+$ channel via the invariant mass distributions corresponded to the favorite resonance state $a_1^+$
at the LHC experiments.
When the relation of the decay rates ${\cal B}(a_1^+ \to \pi^+ \pi^- \pi^+) \approx {\cal B}(a_1^+
\to \pi^+ \pi^0 \pi^0) \sim 50\%$ is adopted~\cite{Aubert:2007kpb}, then the branching fraction
of $B_c^+ \to J/\psi \pi^+ \pi^- \pi^+$ could be derived
under narrow-width approximation as
\beq
{\cal B}(B_c^+ \to J/\psi \pi^+\pi^-\pi^+)_{\rm iPQCD}
&\equiv&
{\cal B}(B_c^+ \to J/\psi a_1^+)\cdot {\cal B}(a_1^+ \to \pi^+\pi^-\pi^+)
=
(2.95^{+0.88}_{-0.70}) \times 10^{-3}\;,
\label{eq:br-psia13p}
\eeq
which is consistent surprisingly well with the predictions using $B_c \to J/\psi$ form factors in three different models~\cite{Likhoded:2009ib} and would be tested by
the experiments at LHC in the future. Subsequently, the ratio $R_{3\pi/\pi}^{\rm Theo}$ between
the branching fractions of $B_c^+ \to J/\psi \pi^+ \pi^- \pi^+$ and $B_c^+ \to J/\psi \pi^+$ could
be obtained straightforwardly as $2.52^{+0.05}_{-0.10}$ in the iPQCD formalism, which is clearly in perfect
consistency with data reported by the CMS and LHCb Collaborations and the Heavy Flavor Averaging Group.
Theoretically, the $B_c^+ \to J/\psi
\pi^+ \pi^- \pi^+$ decay has been investigated in Refs.~\cite{Likhoded:2009ib,Rakitin:2009ya,
Wang:2012vna,Luchinsky:2012rk,Qiao:2012hp} with different ratios $R_{3\pi/\pi}$, however, which are basically
consistent with the current measurements, except for that with the result 1.5~\cite{Rakitin:2009ya}.
The future tests on $R_{a_1/\pi}^{\rm Theo}$ and $R_{3\pi/\pi}^{\rm Theo}$ in the iPQCD
formalism with high precision at the LHC, even CEPC experiments could help understand the property
of $a_1$ meson.

Additionally, with the ratio $R_{2K\pi/\pi}^{\rm Exp}$ in Eq.~(\ref{eq:ex-r2kp-2013-LHCb}) and the branching
fractions of $B_c^+ \to J/\psi a_1^+$ and $B_c^+ \to J/\psi \pi^+$ in the iPQCD formalism, thus
the branching ratio of $a_1^+ \to K^+ K^- \pi^+$ could be deduced under narrow-width approximation
as
\beq
{\cal B}(a_1^+ \to K^+ K^- \pi^+)_{\rm iPQCD} &\equiv&
R_{2K\pi/\pi}^{\rm Exp} \cdot
\frac{{\cal B}(B_c^+ \to J/\psi \pi^+)}{{\cal B}(B_c^+ \to J/\psi a_1^+)}
\approx
(10.5^{+2.0}_{-1.9})\% \;.
\eeq
The detection on this
$a_1^+ \to K^+ K^- \pi^+$ branching ratio would help understand the nature of $a_1$ that is usually
provided from the hadron physics side.
Meanwhile, with the help of $R_{2K\pi/3\pi}^{\rm Exp}$ in Eq.~(\ref{eq:ex-r2kp3p-2021-lhcb}) and ${\cal B}(B_c^+ \to J/\psi \pi^+ \pi^- \pi^+)_{\rm iPQCD}$ in Eq.~(\ref{eq:br-psia13p}), we could derive the branching fraction of $B_c^+ \to J/\psi K^+ K^-\pi^+$ in the iPQCD
formalism as,
\beq
{\cal B}(B_c^+ \to J/\psi K^+ K^- \pi^+)_{\rm iPQCD}
&\equiv&
R_{2K\pi/3\pi}^{\rm Exp}\cdot {\cal B}(B_c^+ \to J/\psi \pi^+ \pi^- \pi^+)_{\rm iPQCD} =
(5.46^{+1.68}_{-1.40}) \times 10^{-4}\;,
\eeq
which is consistent with the predictions given in different form factors~\cite{Luchinsky:2013yla} within a bit large errors.
These two values could be confronted with the future measurements.

In order to help investigate the behavior between the vector $\rho$ meson and the axial-vector $a_1$
and $b_1$ ones, we also define the following two ratios with the $B_c^+ \to J/\psi \rho^+$ decay rate,
\beq
R_{a_1/\rho}^{\rm Theo} &\equiv&
\frac{{\cal B}(B_c^+ \to J/\psi a_1^+)}{{\cal B}(B_c^+ \to J/\psi \rho^+)}
= 1.60^{+0.10}_{-0.11}
%^{+0.00+0.00+0.00+0.10+0.00+0.00}_{-0.01-0.00-0.00-0.11-0.01-0.00}
\;,
\qquad
R_{b_1/\rho}^{\rm Theo} \equiv
\frac{{\cal B}(B_c^+ \to J/\psi b_1^+)}{{\cal B}(B_c^+ \to J/\psi \rho^+)}
= 0.21^{+0.09}_{-0.07}
%^{+0.01+0.00+0.00+0.02+0.09+0.01}_{-0.00-0.00-0.00-0.01-0.07-0.00}
\;.
\eeq
These two ratios are expected to be measured in the near future.

Meantime, the {\it CP}-averaged longitudinal polarization fractions of the $B_c^+ \to J/\psi a_1^+$
and $J/\psi b_1^+$ channels are predicted in the iPQCD formalism theoretically as,
\beq
f_L(B_c^+ \to J/\psi a_1^+)&=&
(74.8^{+0.0}_{-0.3})
%^{+0.0}_{-0.2}(\beta_{B_c})
%^{+}_{-}(f_M)
%^{+0.0}_{-0.0}(f_{B_c})
%^{+0.0}_{-0.0}(f_{J/\psi})
%^{+0.0}_{-0.0}(f_{a_1})
%^{+0.0}_{-0.0}(a_{a_1})
%^{+0.0}_{-0.0}(a_1^t)
%^{+0.0}_{-0.0}(a_2^p)
%^{+0.0}_{-0.2}(a_t)
\% \;,
\qquad
f_L(B_c^+ \to J/\psi b_1^+) =
(98.9^{+0.0}_{-0.0})
%^{+0.0}_{-0.0}(\beta_{B_c})
%^{+}_{-}(f_M)
%^{+0.0}_{-0.0}(f_{B_c})
%^{+0.0}_{-0.0}(f_{J/\psi})
%^{+0.0}_{-0.0}(f_{b_1})
%^{+0.0}_{-0.0}(b_{b_1})
%^{+0.0}_{-0.0}(b_0^p)
%^{+0.0}_{-0.0}(b_1^p)
%^{+0.0}_{-0.0}(b_2^t)
%^{+0.0}_{-0.0}(a_t)
\% \;.
\eeq
It is evident that the decay $B_c^+ \to J/\psi b_1^+$ is governed absolutely by the longitudinal
contributions.

And, for the two $\Delta S =1$ modes $B_c^+ \to J/\psi {K^{(\prime)+}_1}$,
the branching fractions are predicted in the iPQCD formalism with two different mixing angles as follows,
\Rlue{\beq
{\cal B}(B_c^+ \to J/\psi K_1^+)&=&
\left\{ \begin{array}{ll}
2.20^{+0.54}_{-0.43}(\beta_{B_c})
^{+0.23}_{-0.26}(f_M)
%^{+0.04}_{-0.13}(f_{B_c})
%^{+0.15}_{-0.15}(f_{J/\psi}) 
%^{+0.15}_{-0.15}(f_{K_{1A}})
%^{+0.08}_{-0.08}(f_{K_{1B}})
^{+0.61}_{-0.53}(B_{K_1}) 
%^{+0.00}_{-0.00}(a_0^t)
%^{+0.12}_{-0.12}(a_1^p)
%^{+0.00}_{-0.00}(a_1^t)
%^{+0.00}_{-0.00}(a_2^p)
%^{+0.00}_{-0.00}(a_2^t) 
%^{+0.54}_{-0.47}(b_0^p)
%^{+0.26}_{-0.22}(b_1^p)
%^{+0.00}_{-0.00}(b_1^t)
%^{+0.00}_{-0.00}(b_2^p)
%^{+0.00}_{-0.00}(b_2^t) 
\times 10^{-4}
%\;\;\; ({\rm \theta_{K_1}=33^\circ})
& \vspace{0.12cm}
\\
3.51^{+0.90}_{-0.71}(\beta_{B_c})
^{+0.41}_{-0.40}(f_M)
%^{+0.07}_{-0.07 }(f_{B_c})
%^{+0.24}_{-0.24 }(f_{J/\psi}) 
%^{+0.32}_{-0.31 }(f_{K_{1A}})
%^{+0.05 }_{-0.05}(f_{K_{1B}})
^{+0.50}_{-0.46}(B_{K_1}) 
%^{+0.00}_{-0.00}(a_0^t)
%^{+0.18 }_{-0.16}(a_1^p)
%^{+0.00}_{-0.00}(a_1^t)
%^{+0.00}_{-0.00}(a_2^p)
%^{+0.00}_{-0.00}(a_2^t) 
%^{+0.44}_{-0.41 }(b_0^p)
%^{+0.14}_{-0.13}(b_1^p)
%^{+0.00}_{-0.00}(b_1^t)
%^{+0.00}_{-0.00}(b_2^p)
%^{+0.00}_{-0.00}(b_2^t)  
\times 10^{-4}
%\;\;\; ({\rm \theta_{K_1}=58^\circ})
& \end{array} \right.,
\\
{\cal B}(B_c^+ \to J/\psi {K'_1}^+)&=&
\left\{ \begin{array}{ll}
2.11^{+0.66}_{-0.47}(\beta_{B_c})
^{+0.29}_{-0.27}(f_M)
%^{+0.04}_{-0.04}(f_{B_c})
%^{+0.15}_{-0.14}(f_{J/\psi}) 
%^{+0.24}_{-0.23}(f_{K_{1A}})
%^{+0.02}_{-0.02}(f_{K_{1B}})
^{+0.35}_{-0.32}(B_{K_1}) 
%^{+0.00}_{-0.00}(a_0^t)
%^{+0.01}_{-0.00}(a_1^p)
%^{+0.00}_{-0.00}(a_1^t)
%^{+0.00}_{-0.00}(a_2^p)
%^{+0.00}_{-0.00}(a_2^t) 
%^{+0.35}_{-0.32}(b_0^p)
%^{+0.01}_{-0.00}(b_1^p)
%^{+0.00}_{-0.00}(b_1^t)
%^{+0.00}_{-0.00}(b_2^p)
%^{+0.00}_{-0.00}(b_2^t)  
\times 10^{-4}
%\;\;\; ({\rm \theta_{K_1}=33^\circ})
& \vspace{0.12cm}
\\
8.01^{+2.92}_{-2.04}(\beta_{B_c})
^{+0.96}_{-0.92}(f_M)
%^{+0.16}_{-0.17}(f_{B_c})
%^{+0.56}_{-0.55 }(f_{J/\psi}) 
%^{+0.76}_{-0.71 }(f_{K_{1A}})
%^{+0.10}_{-0.09}(f_{K_{1B}})
^{+3.17}_{-2.41}(B_{K_1}) 
%^{+0.00}_{-0.00}(a_0^t)
%^{+0.50}_{-0.43 }(a_1^p)
%^{+0.00}_{-0.00}(a_1^t)
%^{+0.01}_{-0.00}(a_2^p)
%^{+0.00}_{-0.00}(a_2^t) 
%^{+2.91 }_{-2.23}(b_0^p)
%^{+1.15 }_{-0.80}(b_1^p)
%^{+0.00}_{-0.00}(b_1^t)
%^{+0.00}_{-0.00}(b_2^p)
%^{+0.00}_{-0.00}(b_2^t) 
\times 10^{-5}
%\;\;\; ({\rm \theta_{K_1}=58^\circ})
& \end{array} \right.,
\eeq}
where the first (second) entry corresponds to the value obtained at $\theta_{K_1} = 33^\circ\ (58^\circ)$.
The similar patterns also appear in the following observables for related modes. In the numerical results,
the dominant errors arise from the uncertainties of the shape parameter
$\beta_{B_c}$ and from the combined Gegenbauer moment $B_{K_1}$ of $a_{K_{1A}}$ and $a_{K_{1B}}$ (See
Table~\ref{tab:a-vectors} for detail.). For the decay $B_c^+ \to J/\psi K_1^+$, the iPQCD value for
branching fraction at $\theta_{K_1} \sim 33^\circ$ \Rlue{is consistent}
%can compete 
with that at $\theta_{K_1} \sim 58^\circ$,
while, for the decay $B_c^+ \to J/\psi {K'_1}^+$, the result for ${\cal B}(B_c^+ \to J/\psi {K'_1}^+)$
at $\theta_{K_1} \sim 33^\circ$ is roughly \Rlue{four}
%two 
times larger than that at $\theta_{K_1} \sim 58^\circ$. As inferred from Refs.~\cite{ParticleDataGroup:2020ssz,LHCb:2021tdf}, the $B_c^+ \to J/\psi {K'_1}^+$ decay might be explored through the ratio $R_{2\pi K/3\pi}^{\rm Exp} \equiv
{\cal B}(B_c^+ \to J/\psi K^+ \pi^- \pi^+) / {\cal B}(B_c^+ \to J/\psi \pi^+ \pi^- \pi^+)$ in the near future, where the $B_c^+ \to J/\psi K^+ \pi^- \pi^+$ mode proceeds largely via $K^{*0} \to K^+ \pi^-$.
As a by-product, the branching fraction of $B_c^+ \to J/\psi K^+ \pi^- \pi^+$ in the iPQCD formalism could be derived via
the currently measured ratio $R_{2\pi K/3\pi}^{\rm Exp} = (6.4 \pm 1.0) \times 10^{-2}$~\cite{LHCb:2021tdf} and
the iPQCD branching fraction
${\cal B}(B_c^+ \to J/\psi \pi^+ \pi^- \pi^+)$ as the following,
\beq
{\cal B}(B_c^+ \to J/\psi K^+ \pi^- \pi^+)_{\rm iPQCD} &=&
R_{2\pi K/3\pi}^{\rm Exp}\cdot {\cal B}(B_c^+ \to J/\psi \pi^+ \pi^- \pi^+)_{\rm iPQCD}
=(1.89^{+0.63}_{-0.54}) \times 10^{-4}\;,
\eeq
which is in good consistency
with the predictions~\cite{Luchinsky:2013yla} within errors, and is also expected to be detected soon at experiments.

We also predict the longitudinal polarization fractions for the decays $B_c^+ \to J/\psi K_1^{(\prime)+}$
in the iPQCD formalism under two referenced angles as follows,
\Rlue{\beq
f_L(B_c^+ \to J/\psi K_1^+)&=&
\left\{ \begin{array}{ll}
(82.0^{+2.8}_{-3.0})
%%^{+0.0}_{-0.2}(\beta_{B_c})
%^{+}_{-}(f_M)
%%^{+0.0}_{-0.0}(f_{B_c})
%%^{+0.0}_{-0.0}(f_{J/\psi})
%%^{+0.4}_{-0.4}(f_{K_{1A}})
%%^{+0.4}_{-0.5}(f_{K_{1B}}) 
%^{+0.0}_{-0.0}(a_0^t)
%^{+1.0}_{-1.2}(a_1^p)
%^{+0.0}_{-0.0}(a_1^t)
%^{+0.1}_{-0.0}(a_2^p)
%^{+0.0}_{-0.0}(a_2^t) 
%^{+1.4}_{-1.2}(b_0^p)
%^{+2.1}_{-2.4}(b_1^p)
%^{+0.0}_{-0.0}(b_1^t)
%^{+0.0}_{-0.0}(b_2^p)
%^{+0.0}_{-0.0}(b_2^t) 
\%
%\;\;\;\;\ ({\rm \theta_{K_1}=33^\circ})
& \vspace{0.12cm}
\\
(76.0^{+1.8}_{-1.7})
%%^{+0.0}_{-0.2}(\beta_{B_c})
%^{+}_{-}(f_M)
%%^{+0.0}_{-0.0}(f_{B_c})
%%^{+0.0}_{-0.0}(f_{J/\psi})
%%^{+0.3 }_{-0.2}(f_{K_{1A}})
%%^{+0.2}_{-0.2}(f_{K_{1B}}) 
%^{+0.0}_{-0.0}(a_0^t)
%^{+1.4 }_{-1.3}(a_1^p)
%^{+0.0}_{-0.0}(a_1^t)
%^{+0.0}_{-0.0}(a_2^p)
%^{+0.0}_{-0.0}(a_2^t) 
%^{+0.3}_{-0.2}(b_0^p)
%^{+1.1 }_{-1.1}(b_1^p)
%^{+0.0}_{-0.0}(b_1^t)
%^{+0.0}_{-0.0}(b_2^p)
%^{+0.0}_{-0.0}(b_2^t) 
\%
%\;\;\;\;\ ({\rm \theta_{K_1}=58^\circ})
& \end{array} \right.\;,
\qquad
f_L(B_c^+ \to J/\psi {K'_1}^+)=
\left\{ \begin{array}{ll}
(67.7^{+1.1}_{-0.7})
%%^{+0.7}_{-0.6}(\beta_{B_c})
%^{+}_{-}(f_M)
%%^{+0.0}_{-0.0}(f_{B_c})
%%^{+0.0}_{-0.0}(f_{J/\psi})
%%^{+0.1}_{-0.1}(f_{K_{1A}})
%%^{+0.1}_{-0.1}(f_{K_{1B}}) 
%^{+0.0}_{-0.0}(a_0^t)
%^{+0.6}_{-0.2}(a_1^p)
%^{+0.0}_{-0.0}(a_1^t)
%^{+0.0}_{-0.0}(a_2^p)
%^{+0.0}_{-0.0}(a_2^t) 
%^{+0.3}_{-0.3}(b_0^p)
%^{+0.5}_{-0.2}(b_1^p)
%^{+0.0}_{-0.0}(b_1^t)
%^{+0.0}_{-0.0}(b_2^p)
%^{+0.0}_{-0.0}(b_2^t) 
\%
%\;\;\;\;\ ({\rm \theta_{K_1}=33^\circ})
& \vspace{0.12cm}
\\
(70.6^{+4.7}_{-3.6})
%%^{+1.4}_{-1.6}(\beta_{B_c})
%^{+}_{-}(f_M)
%%^{+0.0}_{-0.0}(f_{B_c})
%%^{+0.0}_{-0.0}(f_{J/\psi})
%%^{+ 0.7}_{-0.6}(f_{K_{1A}})
%%^{+0.7}_{-0.7}(f_{K_{1B}}) 
%^{+0.0}_{-0.0}(a_0^t)
%^{+1.3}_{- 1.3}(a_1^p)
%^{+0.0}_{-0.0}(a_1^t)
%^{+0.0}_{-0.0}(a_2^p)
%^{+0.0}_{-0.0}(a_2^t) 
%^{+3.1}_{- 1.6}(b_0^p)
%^{+2.8 }_{-2.3}(b_1^p)
%^{+0.0}_{-0.0}(b_1^t)
%^{+0.0}_{-0.0}(b_2^p)
%^{+0.0}_{-0.0}(b_2^t) 
\%
%\;\;\;\;\ ({\rm \theta_{K_1}=58^\circ})
& \end{array} \right..
\eeq}
It is clear to observe that, within the theoretical errors, the longitudinal polarization decay
amplitudes dominate the $B_c^+ \to J/\psi K_1^{(\prime)+}$ modes.
%, however, both the longitudinal
%and the transverse polarization decay amplitudes generally compete with each other in
%the $B_c^+ \to J/\psi {K'_1}^+$ channel. 
It means that, generally speaking, the slightly 
%significantly 
constructive (destructive)
interferences occur at the longitudinal polarization in the $B_c^+ \to J/\psi K_1^+$ 
($B_c^+ \to J/\psi K_1^{\prime +}$) mode. The future
measurements on these two decays might reveal the information
of the mixing angle $\theta_{K_1}$ between $K_1^+$ and ${K'_1}^+$.

Meanwhile, for the convenience of future probes to the decays $B_c^+ \to J/\psi K_1^+$ and $J/\psi
{K'_1}^+$, the relative ratios of the branching fractions between
$B_c^+ \to J/\psi K_1^{(\prime)+}$
and $B_c^+ \to J/\psi \pi^+$ could be derived in the  iPQCD formalism as follows,
\Rlue{\beq
R^{\rm Theo}_{K_1/\pi}&\equiv&
\frac{{\cal B}(B_c^+ \to J/\psi K_1^+)}{{\cal B}(B_c^+ \to J/\psi \pi^+)}=
\left\{ \begin{array}{ll}
0.19^{+0.05}_{-0.05} 
& \vspace{0.12cm} \\
0.30^{+0.04}_{-0.04} 
&  \\ \end{array} \right.,
\quad
R^{\rm Theo}_{K'_1/\pi} \equiv
\frac{{\cal B}(B_c \to J/\psi {K'_1}^{+})}{{\cal B}(B_c \to J/\psi \pi^+)}=
\left\{ \begin{array}{ll}
0.18^{+0.03}_{-0.03} 
& \vspace{0.12cm} \\
0.07^{+0.03}_{-0.02} 
&  \\ \end{array} \right..
\eeq}
We also present the ratios between the $\Delta S =1$ and $\Delta S=0$ decay rates in these $B_c^+
\to J/\psi A^+$ modes as the following,
\Rlue{\beq
R^{\rm Theo}_{K_1/a_1}&\equiv&
\frac{{\cal B}(B_c^+ \to J/\psi K_1^+)}{{\cal B}(B_c^+ \to J/\psi a_1^+)}=
\left\{ \begin{array}{ll} 
0.04^{+0.01}_{-0.01} 
& \vspace{0.12cm} \\ 
0.06^{+0.01}_{-0.01} 
&  \\ \end{array} \right.,
\quad
R^{\rm Theo}_{K'_1/a_1}\equiv
\frac{{\cal B}(B_c^+ \to J/\psi {K'_1}^+)}{{\cal B}(B_c^+ \to J/\psi a_1^+)}=
\left\{ \begin{array}{ll} 
0.04^{+0.00}_{-0.01} 
& \vspace{0.12cm} \\  
0.01^{+0.01}_{-0.00} 
&  \\ \end{array} \right.\;,
\eeq}
and
\Rlue{\beq
R^{\rm Theo}_{K_1/b_1}&\equiv&
\frac{{\cal B}(B_c^+ \to J/\psi K_1^+)}{{\cal B}(B_c^+ \to J/\psi b_1^+)}=
\left\{ \begin{array}{ll} 
0.28^{+0.03}_{-0.04} 
& \vspace{0.12cm} \\ 
0.44^{+0.13}_{-0.08} 
&  \\ \end{array} \right.,
\quad
R^{\rm Theo}_{K'_1/b_1}\equiv
\frac{{\cal B}(B_c^+ \to J/\psi {K'_1}^+)}{{\cal B}(B_c^+ \to J/\psi b_1^+)}=
\left\{ \begin{array}{ll} 
0.27^{+0.07}_{-0.05} 
& \vspace{0.12cm}
\\ 
0.10^{+0.01}_{-0.00} 
&  \\ \end{array} \right. \;.
\eeq}
These ratios would be helpful to explore the QCD dynamics in the considered axial-vector mesons,
especially in $K_1$ and ${K'_1}$.

\subsection{\boldmath $B_c^+ \to J/\psi S^+$}
\label{ssec:psis}

As we currently known, our colleagues categorize the light scalars in the two-quark structure
into two different scenarios~\cite{Cheng:2005nb},
namely, the scalars below 1~GeV could be considered as the $q\bar q$-ground states and those
around 1.5~GeV could be viewed as the first excited states correspondingly in scenario 1 (S1),
while the scalars around 1.5~GeV might be the lowest-lying $q\bar q$-bound states and those
below 1~GeV have to be the four-quark states in scenario 2 (S2). This means explicitly that
$a_0(980)^+$ (It will be denoted as $a_0^+$ for convenience.) and $\kappa^+$ in S1 and $a_0(1450)^+$
and ${K_0^*(1430)}^+$ (They will be expressed as ${a'_0}^+$ and ${K_0^*}^+$ for simplicity.) in
both S1 and S2 will be taken into this study, due to the availability of factorization approach.
Different from the factorizable-emission-diagrams-dominated decays
$B_c^+ \to J/\psi (P^+, V^+)$, due to the highly small vector decay constant $f_{S}$, the
factorizable emission diagrams are strongly suppressed in the $B_c^+ \to J/\psi S^+$ modes.
%%%========================================================================================================
\begin{table}[htb]
\caption{
The decay amplitudes (in units of $10^{-3}$GeV$^{-3}$) from different diagrams
of $B_c^+ \to J/\psi S^+$ in the iPQCD formalism.
The upper (lower) entry corresponds to the scalars ${a'_0}^+$ and ${K_0^*}^+$
in scenario 1 (2) at every line. For comparison, the decay amplitudes of
$B_c^+ \to J/\psi \pi^+$ are also provided in the last column. For the sake
of simplicity, only the central values are quoted for clarifications.}
\label{tab:DecAmps}
\begin{center}\vspace{-0.5cm}{%\footnotesize
\begin{tabular}[t]{c||c|c|c|c||c}
\hline  \hline
Modes   & $B_c^+ \to J/\psi a_0^+$  & $B_c^+ \to J/\psi \kappa^+$
& $B_c^+ \to J/\psi {a'_0}^+$ & $B_c^+ \to J/\psi {K_0^*}^+$
& $B_c \to J/\psi \pi^+$ \\
\hline \hline
Decay Amplitudes ($fe$)
&$ - 0.11 - {\it i} 0.73 $
&$ - 1.17 - {\it i} 7.60 $
&$\begin{array}{cc}
0.09 + {\it i} 0.63
\\
- 0.15 -{\it i} 0.97
\end{array}$
&$\begin{array}{cc}
1.07 + {\it i} 7.06
\\
- 1.60 -{\it i} 10.60
\end{array}$
&$18.21 + {\it i} 91.88$
\\
\hline
Decay Amplitudes ($nfe$)
&$- 48.13 + {\it i} 52.70$
&$- 7.87 + {\it i} 12.70$
& $\begin{array}{cc}
- 68.32 + {\it i} 33.43
\\
- 44.16 +{\it i} 7.08
\end{array}$
&$\begin{array}{cc}
- 11.39 + {\it i} 6.47
\\
- 10.30 +{\it i} 2.90
\end{array}$
&$- 2.04 + {\it i} 3.22$
\\
\hline \hline
\end{tabular}}
\end{center}
\end{table}
%%%%========================================================================================================
We present explicitly the decay amplitudes of $B_c^+ \to J/\psi S^+$ from
the factorizable emission and nonfactorizable emission diagrams respectively in Table~\ref{tab:DecAmps}~\footnote{
Due to the SU(3) flavor symmetry breaking effects, specifically,
the strange quark mass and the up or down quark mass satisfying the relation of
$m_s \gg m_{u,d}$, then the considerable contributions
induced by the vector decay constants $f_{\kappa^+}$ and $f_{{K_0^*}^+}$ according
to Eq.~(\ref{eq:vd-sd}) appear
from the factorizable emission diagrams in the $B_c^+ \to J/\psi S^+ (\Delta S =1)$
channels, relative to those in the $B_c^+ \to J/\psi S^+
(\Delta S =0)$ modes with tiny isospin symmetry breaking. }.
Therefore, the estimations going beyond naive factorization are essential for us to find out the
useful constraints on $\beta_{B_c}$ with the help of large branching fractions arising almost from
the nonfactorizable emission contributions. Moreover,
different from the decay $B_c^+ \to J/\psi b_1^+$ with angular decomposition,
the decays $B_c^+ \to J/\psi  S^+$ have only longitudinal contributions because of conservation
of the angular momentum.

The {\it CP}-averaged branching fractions of the $B_c^+ \to J/\psi S^+$ decays in the iPQCD formalism
are given as,
\beq
{\cal B}(B_c^+ \to J/\psi a_0^+)&=&
5.98^{+1.77}_{-1.39}(\beta_{B_c})
^{+0.80}_{-0.78}(f_M)
%^{+0.12}_{-0.12}(f_{B_c})
%^{+0.42}_{-0.41}(f_{J/\psi})
%^{+0.67}_{-0.65}(\bar f_{a_0})
^{+1.31}_{-1.22}(B_i)
%^{+1.30}_{-1.21}(B_1)
%^{+0.15}_{-0.15}(B_3)
%^{+0.00}_{-0.03}(a_t)
\times 10^{-4}\;,
\\
{\cal B}(B_c^+ \to J/\psi \kappa^+)&=&
1.31^{+0.50}_{-0.36}(\beta_{B_c})
^{+0.18}_{-0.18}(f_M)
%^{+0.02}_{-0.03}(f_{B_c})
%^{+0.09}_{-0.09}(f_{J/\psi})
%^{+0.16}_{-0.15}(\bar f_{\kappa})
^{+0.41}_{-0.35}(B_i)
%^{+0.41}_{-0.35}(B_1)
%^{+0.02}_{-0.04}(B_3)
%^{+0.05}_{-0.02}(a_t)
\times 10^{-5}\;,
\eeq
and
\beq
{\cal B}(B_c^+ \to J/\psi {a'_0}^+) &=&
\left\{ \begin{array}{ll}
6.39^{+2.09}_{-1.56}(\beta_{B_c})
^{+1.52}_{-1.37}(f_M)
%^{+0.14}_{-0.13}(f_{B_c})
%^{+0.45}_{-0.43}(f_{J/\psi})
%^{+1.44}_{-1.29}(\bar f_{a'_0})
^{+2.19}_{-1.89}(B_i)
%^{+2.13}_{-1.82}(B_1)
%^{+0.53}_{-0.50}(B_3)
%^{+0.04}_{-0.00}(a_t)
\times 10^{-4}
%\;\;\;\;\ ({\rm S1})
&
\vspace{0.12cm}
\\
2.20^{+0.71}_{-0.54}(\beta_{B_c})
^{+0.54}_{-0.48}(f_M)
%^{+0.05}_{-0.04}(f_{B_c})
%^{+0.16}_{-0.15}(f_{J/\psi})
%^{+0.51}_{-0.45}(\bar f_{a'_0})
^{+1.37}_{-1.02}(B_i)
%^{+1.32}_{-0.97}(B_1)
%^{+0.38}_{-0.30}(B_3)
%^{+0.11}_{-0.11}(a_t)
\times 10^{-4}
%\;\;\;\;\;({\rm S2})
&  \\ \end{array} \right.,
\\
{\cal B}(B_c^+ \to J/\psi K_0^{*+}) &=&
\left\{ \begin{array}{ll}
3.22^{+0.81}_{-0.66}(\beta_{B_c})
^{+0.74}_{-0.64}(f_M)
%^{+0.06}_{-0.07}(f_{B_c})
%^{+0.22}_{-0.22}(f_{J/\psi})
%^{+0.70}_{-0.60}(\bar f_{K_0^*})
^{+0.37}_{-0.35}(B_i)
%^{+0.35}_{-0.33}(B_1)
%^{+0.13}_{-0.12}(B_3)
%^{+0.07}_{-0.07}(a_t)
\times 10^{-5}
%\;\;\;\;\ ({\rm S1})
& \vspace{0.12cm}
\\
2.23^{+0.87}_{-0.70}(\beta_{B_c})
^{+0.55}_{-0.51}(f_M)
%^{+0.05}_{-0.04}(f_{B_c})
%^{+0.16}_{-0.15}(f_{J/\psi})
%^{+0.52}_{-0.49}(\bar f_{K_0^*})
^{+0.65}_{-0.45}(B_i)
%^{+0.64}_{-0.45}(B_1)
%^{+0.10}_{-0.05}(B_3)
%^{+0.05}_{-0.00}(a_t)
\times 10^{-5}
%\;\;\;\;\;({\rm S2})
&  \\ \end{array} \right.,
\eeq
where the first (second) entry corresponds to the value obtained in S1 (S2). The similar patterns
also appear in the following ratios for related modes. We stress that the less experimental
constraints on the shape parameter $\beta_{B_c}$ of $B_c$ meson, and on the scalar decay constant
$\bar f_{S}$ and the Gegenbauer moments $B_{i}$ in
the light-cone distribution amplitudes of scalars result in the remarkably large errors in theory.

Generally speaking, the theoretical uncertainties induced by the input parameters are usually canceled
to a great extent in the relative ratios of the branching fractions. The ratios between the
corresponding branching fractions of the
$\Delta S=1$ and $\Delta S=0$ modes in the $B_c^+ \to J/\psi S^+$ decays can then be read as,
\beq
R^{\rm Theo}_{\kappa/a_0}&\equiv&
\frac{{\cal B}(B_c^+ \to J/\psi \kappa^+)}{{\cal B}(B_c^+ \to  J/\psi a_0^+)}=
%0.02^{+0.00}_{-0.00}
0.022^{+0.002}_{-0.002}
%^{+0.001+0.000+0.000+0.000+0.002+0.001}_{-0.001-0.000-0.000-0.000-0.002-0.000} \times 10^{-2}
\;,
%;
\quad
R^{\rm Theo}_{K_0^*/a'_0} \equiv
\frac{{\cal B}(B_c^+ \to J/\psi K_0^{*+})}{{\cal B}(B_c^+ \to J/\psi {a'_0}^+)}=\left\{ \begin{array}{ll}
%0.05^{+0.01}_{-0.01}
0.050^{+0.014}_{-0.008}
%^{+0.003+0.000+0.000+0.001+0.014+0.001}_{-0.002-0.000-0.000-0.000-0.008-0.001} \times 10^{-2}
%\;\;\;\;\ ({\rm S1})
& \vspace{0.12cm} \\
%0.10^{+0.05}_{-0.02}
0.101^{+0.050}_{-0.022}
%^{+0.005+0.000+0.000+0.000+0.049+0.006}_{-0.009-0.000-0.000-0.002-0.020-0.003} \times 10^{-2}
%\;\;\;\;\;({\rm S2})
&  \\ \end{array} \right..
\eeq
And, the relative ratios of ${\cal B}(B_c^+ \to J/\psi S^+)$ over ${\cal B}(B_c^+ \to J/\psi \pi^+)$
are presented for future detections as follows,
\beq
R^{\rm Theo}_{a_0/\pi}&\equiv&
\frac{{\cal B}(B_c^+ \to J/\psi a_0^+)}{{\cal B}(B_c^+ \to J/\psi \pi^+)}=
0.51^{+0.13}_{-0.12}
%0.511^{+0.126}_{-0.120}
%^{+0.012+0.002+0.001+0.057+0.112+0.002}_{-0.023-0.001-0.000-0.055-0.104-0.004} \times 10^{0}
\;,
\quad
R^{\rm Theo}_{\kappa/\pi} \equiv
\frac{{\cal B}(B_c^+ \to J/\psi \kappa^+)}{{\cal B}(B_c^+ \to J/\psi \pi^+)}=
%0.01^{+0.01}_{-0.00}
0.011^{+0.005}_{-0.004}
%^{+0.001+0.000+0.000+0.002+0.004+0.001}_{-0.001-0.000-0.000-0.002-0.003-0.000} \times 10^{-2}
\;,
\eeq
and
\beq
R^{\rm Theo}_{a'_0/\pi} \equiv
\frac{{\cal B}(B_c^+ \to J/\psi {a'_0}^+)}{{\cal B}(B_c^+ \to J/\psi \pi^+)}
&=&
\left\{ \begin{array}{ll}
0.55^{+0.19}_{-0.17}
%0.546^{+0.189}_{-0.172}
%^{+0.027+0.003+0.001+0.187+0.005}_{-0.032-0.002-0.000-0.161-0.001} \times 10^{0}
%\;\;\;\;\ ({\rm S1})
& \vspace{0.12cm} \\
0.19^{+0.12}_{-0.09}
%0.188^{+0.117}_{-0.088}
%^{+0.009+0.001+0.001+0.117+0.008}_{-0.011-0.000-0.000-0.087-0.008} \times 10^{-1}
%\;\;\;\;\;({\rm S2})
&  \\ \end{array} \right.,
\quad
R^{\rm Theo}_{K_0^*/\pi} \equiv
\frac{{\cal B}(B_c^+ \to J/\psi {K_0}^{*+})}{{\cal B}(B_c^+ \to J/\psi \pi^+)}
=
\left\{ \begin{array}{ll}
%0.03^{+0.00}_{-0.00}
0.028^{+0.003}_{-0.003}
%^{+0.000+0.000+0.000+0.003+0.000}_{-0.001-0.001-0.000-0.003-0.001} \times 10^{-2}
%\;\;\;\;\ ({\rm S1})
& \vspace{0.12cm} \\
%0.02^{+0.01}_{-0.01}
0.019^{+0.006}_{-0.005}
%^{+0.002+0.000+0.000+0.006+0.000}_{-0.003-0.000-0.000-0.004-0.000} \times 10^{-2}
%\;\;\;\;\;({\rm S2})
&  \\ \end{array} \right..
\eeq
Strictly speaking, the errors induced by the hadronic parameters of light scalars
are hard to be effectively canceled due to their unknown nature.
It is found that the errors are still large in the above ratios generally.
Nevertheless, these decay modes can provide chances to understand the QCD dynamics
because they must be studied in the factorization framework of QCD
going beyond naive factorization hypothesis.

\subsection{\boldmath $B_c^+ \to J/\psi T^+$}
\label{ssec:psit}
As discussed in footnote~\ref{fnt:footnote1}, the factorization formulas analogous to the $B_c^+ \to
J/\psi V^+$ decays in the modes $B_c^+ \to J/\psi T^+$ could then be easily got because of
conservation of the angular momentum. But, extremely different from the $B_c^+ \to J/\psi V^+$
decays, the tensor mesons cannot be produced via the vector current. Hence, the
factorizable contributions associated with $T$-emission in these $B_c^+ \to J/\psi T^+$
modes are forbidden intuitively. It means that the decays $B_c^+ \to J/\psi T^+$
must be explored beyond the naive factorization approach. Their branching fractions
are contributed completely by the nonfactorizable emission diagrams. It is emphasized that,
due to few studies on the light-cone distribution amplitudes of tensor meson, only the available
asymptotic forms of tensor meson's distribution amplitudes are adopted tentatively in this work.

Therefore, the branching fractions of the $B_c^+ \to  J/\psi T^+$ decays predicted in the iPQCD formalism
are presented as follows,
\beq
{\cal B}(B_c^+ \to J/\psi a_2^+)&=&
1.39^{+0.45}_{-0.33}(\beta_{B_c})
^{+0.19}_{-0.18}(f_M)
%^{+0.03}_{-0.03}(f_{B_c})
%^{+0.10}_{-0.09}(f_{J/\psi})
%^{+0.16}_{-0.15}(f_{a_2})
%^{+0.00}_{-0.00}(f_{a_2}^T)
%^{+0.03}_{-0.02}(a_t)
\times 10^{-4} \;,
\\
{\cal B}(B_c^+ \to J/\psi K_2^{*+})&=&
9.05^{+2.91}_{-2.22}(\beta_{B_c})
^{+1.03}_{-0.98}(f_M)
%^{+0.19}_{-0.18}(f_{B_c})
%^{+0.64}_{-0.61}(f_{J/\psi})
%^{+0.79}_{-0.75}(f_{K^*_2})
%^{+0.00}_{-0.00}(f_{K^*_2}^T)
%^{+0. }_{-0. }(a_\rho)
%^{+0.}_{-0.}(a_\rho)
%^{+0.18}_{-0.26}(a_t)
\times 10^{-6} \;,
\eeq
where the dominant errors come from the shape parameter $\beta_{B_c}$ and the less constrained
decay constant $f_T$, respectively. Based on the assumptions of ${\cal B}(a_2^+ \to \pi^+ \pi^- \pi^+)
\approx {\cal B}(a_2^+ \to \pi^+ \pi^0 \pi^0)$ and the validity of narrow-width approximation associated
with the branching fractions
${\cal B}(a_2 \to 3\pi) = (70.1 \pm 2.7)\%$~\cite{ParticleDataGroup:2020ssz} and ${\cal B}(B_c^+ \to J/\psi a_2^+)
= (1.39^{+0.49}_{-0.38}) \times 10^{-4}$,
the large branching fraction ${\cal B}(B_c^+ \to J/\psi a_2^+ (\to \pi^+ \pi^-\pi^+))_{\rm iPQCD}
= (0.49^{+0.17}_{-0.14}) \times 10^{-4}$ will be tested at the relevant experiments in the near
future. The measurements on this value will help examine the reliability of iPQCD formalism
and further obtain the information of $\beta_{B_c}$, even the tensor mesons' QCD behavior
from the related observables.

The longitudinal polarization fractions of the $B_c^+ \to J/\psi T^+$ modes are also predicted in
the iPQCD formalism as the following,
\beq
f_L(B_c^+ \to J/\psi a_2^+)&=&
(96.1^{+0.1}_{-0.1})
%^{+0.1}_{-0.1}(\beta_{B_c})
%^{+0.0 }_{-0.0 }(f_M)
%%^{+0.0}_{-0.0}(f_{B_c})
%%^{+0.0}_{-0.0}(f_{J/\psi})
%%^{+0.0}_{-0.0}(f_{a_2})
%%^{+0.0}_{-0.0}(f_{a_2}^T)
%^{+0.1}_{-0.0}(a_t)
\% \;,
\qquad
f_L(B_c^+ \to J/\psi K_2^{*+}) =
(95.3^{+0.0}_{-0.2})
%^{+0.1}_{-0.2}(\beta_{B_c})
%^{+0.0 }_{-0.0 }(f_M)
%%^{+0.0}_{-0.0}(f_{B_c})
%%^{+0.0}_{-0.0}(f_{J/\psi})
%%^{+0.0}_{-0.0}(f_{K^*_2})
%%^{+0.0}_{-0.0}(f_{K^*_2}^T)
%^{+0.1}_{-0.0}(a_t)
\%\;,
\eeq
which meets the naively expected hierarchy, i.e., $f_{L} \sim 1$, in the tree-level $\bar b \to
\bar c$ transitions based on quark-helicity conservation~\cite{Ali:1978kn,Suzuki:2001za}.

Like $R_{K/\pi}$ in the $B_c^+ \to J/\psi P^+$ sector, we can define $R_{K_2^*/a_2}$ in the
$B_c^+ \to J/\psi T^+$ decays and predict its value within the iPQCD formalism as follows,
\beq
R^{\rm Theo}_{K_2^*/a_2}&\equiv&
\frac{{\cal B}(B_c^+ \to J/\psi K_2^{*+})}{{\cal B}(B_c^+ \to J/\psi a_2^+)}=
0.065^{+0.002}_{-0.002}
%^{+0.000+0.000+0.000+0.002+0.000}_{-0.001-0.000-0.000-0.002-0.001}
%\times 10^{-2}
\;.
\eeq
It is of great interest to find that this result agrees so well with
$R_{K^*/\rho} \approx |f_{K^*}/f_{\rho}|^2\cdot|V_{us}/V_{ud}|^2$ naively
anticipated in factorization ansatz, and is also very close to the ratio $R_{K/\pi}^{\rm Theo}$.
The underlying reason is that, relative to the predominant
contributions from factorizable emission diagrams while with negligible nonfactorizable emission
contributions in $B_c^+ \to J/\psi V^+$, the $B_c^+ \to J/\psi T^+$ decays are absolutely
contributed from the nonfactorizable emission ones. Therefore, the ratio in the latter decays
could be cleanly written as $|f_{K_2^*}/f_{a_2}|^2\cdot|V_{us}/V_{ud}|^2$ due to the SU(3)-flavor
symmetry in the leading-twist distribution amplitude $\phi_T(x)$, however, that in the former
decays could only be approximately expressed as $|f_{K^*}/f_{\rho}|^2\cdot|V_{us}/V_{ud}|^2$
due to a bit destructive interferences from the nonfactorizable emission decay amplitudes in fact.

The ratios of ${\cal B}(B_c^+ \to J/\psi T^+)$ over ${\cal B}(B_c^+ \to J/\psi \pi^+)$
in the iPQCD formalism could be written as,
\beq
R^{\rm Theo}_{a_2/\pi}&\equiv&
\frac{{\cal B}(B_c^+ \to J/\psi a_2^+)}{{\cal B}(B_c^+ \to J/\psi \pi^+)}=
0.12^{+0.01}_{-0.01}
%0.119^{+0.005}_{-0.006}
%^{+0.005+0.000+0.000+0.001}_{-0.006-0.001-0.000-0.001}
%\times 10^{-1}
\;,
\qquad
R^{\rm Theo}_{K_2^*/\pi} \equiv
\frac{{\cal B}(B_c^+ \to J/\psi K_2^{*+})}{{\cal B}(B_c^+ \to J/\psi \pi^+)}=
%0.01^{+0.00}_{-0.00}
0.008^{+0.000}_{-0.001}
%^{+0.000+0.000+0.000+0.000}_{-0.001-0.000-0.000-0.000} \times 10^{-2}
\;,
\eeq
which will be utilized to help explore these two $B_c^+ \to J/\psi T^+$ decay modes at LHC,
even CEPC experiments in the future.

Because of no very rigorous constraints on the shape parameter $\beta_{B_c}$ in the
$B_c$-meson distribution amplitude and on the Gegenbauer moments in the light-cone distribution
amplitudes of $p$-wave light hadrons from the aspects of current experiments,
we suggest our experimental colleagues to make much more relevant measurements on the predictions,
especially the relative ratios of the branching fractions, to further understand the
involved perturbative and nonperturbative
QCD dynamics in the related channels, despite no easy detections on the individual decay rates.
The related measurements could also help differentiate the reliability of the adopted
approaches and/or methods.

Finally, one more comment is that, within the framework of this iPQCD formalism in association
with the newly derived Sudakov factor by including the charm quark mass effects, we could extend
the related studies to the decays involving $B_c$ to charmonia transitions, such as
$B_c^+ \to (\eta_c, \chi_{cJ}(J=0,1,2), \ldots) M^+$, which request detailed investigations
on the related modes because of the involved complicated dynamics. These studies will be performed
in the future and the numerical results will be presented elsewhere.

\bigskip

%%%--=================================================================
%%%=====            Conclusions&Summary     ==========================
%%5===================================================================
%\section{Conclusions and Summary} \label{sec:summary}

In summary, we have systematically studied the $B_c$-meson decays into $J/\psi$ plus
a light meson by completely including the charm quark mass effects in the iPQCD formalism
at leading order. Several interesting predictions
for the observables such as branching fractions, relative ratios, and longitudinal
polarization fractions are presented explicitly. Based on the numerical results and
phenomenological analyses, we find that
\begin{itemize}
\item
The iPQCD predictions for the ratios between the branching fractions of
$B_c^+ \to J/\psi K^+/ B_c^+ \to J/\psi a_1^+( \to \pi^+ \pi^-\pi^+)$
and $B_c^+ \to J/\psi \pi^+$ are highly consistent with the current data,
though their individual decay rates are not yet available presently.
The branching fractions of CKM-favored
$B_c^+ \to J/\psi P^+$, $B_c^+ \to J/\psi V^+$, and $B_c^+ \to J/\psi a_1^+$
decays, which are predominated by the factorizable emission diagrams,
around ${\cal O}(10^{-3})$ are predicted in the iPQCD formalism and
await near future tests at the LHC experiments.

\item
The iPQCD predictions for the branching fractions of the factorizable-emission-diagrams-suppressed/forbidden
modes such as $B_c^+ \to J/\psi b_1^+$,
$B_c^+ \to J/\psi S^+$, and $B_c^+ \to J/\psi T^+$
are provided theoretically for the first time in the literature and will be confronted with the future
measurements at LHC, even CEPC experiments. Objectively speaking, the investigations on these
decay modes with large nonfactorizable decay amplitudes should go beyond naive
factorization to further help understand the involved
perturbative and nonperturbative QCD dynamics. Phenomenologically,
the precise measurements on this type of decays
could help constrain the shape parameter $\beta_{B_c}$ in $\phi_{B_c}(x)$,
as well as explore the nature of light hadrons.

\item
Almost all of the $B_c^+ \to J/\psi M^+$ decays with polarization contributions are governed
by the longitudinal decay amplitudes, which are consistent with the naive expectation $f_L \sim 1$
in the tree-level $\bar b \to \bar c$ transitions, except for the $B_c^+ \to J/\psi K_1(1400)^+$
channel with possibly destructive interferences between $B_c^+ \to J/\psi K_{1A}^+$ and
$B_c^+ \to J/\psi K_{1B}^+$ in longitudinal polarization. The related iPQCD predictions
will be examined at relevant experiments in the future.

\item
The model-independent ratio between the branching fractions of CKM-suppressed and CKM-favored $B_c^+ \to J/\psi T^+$
decays is obtained and expected to shed light on the information of $\beta_{B_c}$ promisingly.
By utilizing the golden channel $B_c^+ \to J/\psi \pi^+$ as normalization, the relative ratios
between the branching fractions of $B_c^+ \to J/\psi M^+\; (M = V, A, S, T)$ and $B_c^+ \to J/\psi \pi^+$ are
predicted in the iPQCD formalism and would be helpful to search for these decay modes in near future examinations.

\end{itemize}

%%%--=================================================================
%%%=====            Acknowledgements        ==========================
%%5===================================================================

\begin{acknowledgments}
X.L. thanks Profs. H.-n.~Li and Z.J.~Xiao for their useful discussions and
constant encouragements, and also thanks Prof. H.-n.~Li for reading the manuscript
and Ivan Belyaev for a valuable suggestion.
This work is supported by the National Natural Science
Foundation of China under Grants No.~11875033 and No.~11205072,
by the Qing Lan Project of Jiangsu Province(No.~9212218405),
and by the Research Fund of Jiangsu Normal University under Grant No.~HB2016004.
\end{acknowledgments}

%%%%%%%%%%%%%%%%%%%%%%%%%%%%%%%%%%%%%%%%%%%%%%%%%%%%%%%%%%%%%%%%%%%%%%%%%%%%%%%%%%
%                                        Appendix
%%%%%%%%%%%%%%%%%%%%%%%%%%%%%%%%%%%%%%%%%%%%%%%%%%%%%%%%%%%%%%%%%%%%%%%%%%%%%%%%5

\begin{appendix}

\section{ Mesons' distribution amplitudes  }
\label{app:Dis-Amps}

As aforementioned, the distribution amplitudes of initial and final mesons have been
presented in the literature. For the sake of simplicity, we here just collect them
in this Appendix.

For the initial $B_c$ meson, the distribution amplitude $\phi_{B_c}(x, {\bf b})$ in
the conjugate ${\bf b}$ space of transverse momentum ${\bf k}_T$ could be written as
follows~\cite{Liu:2018kuo},
\beq
\phi_{B_c}(x,{\bf b}) &=&   \frac{f_{B_c}}{2\sqrt{2 N_c } }
N_{B_c}x(1-x)\exp\left[-\frac{(1-x)m_c^2+xm_b^2}
{8\beta_{B_c}^2x(1-x)}\right]\exp\left[-2\beta_{B_c}^2x(1-x){\bf b}^2\right]\;,
\eeq
The shape parameter $\beta_{B_c} = 1.0 \pm 0.1$~GeV could enable the distribution of
$\phi_{B_c}(x)$ to coincide with those proposed in Refs.~\cite{Wang:2017bgv,Sun:2015exa}.
The normalization constant $N_{B_c}$
is fixed by the following relation,
\beq
\int_0^1 \phi_{B_c}(x, {\bf b}=0) dx \equiv\int_0^1 \phi_{B_c}(x) dx
&=& \frac{f_{B_c}}{2\sqrt{2 N_c} }\;,
\eeq
where the decay constant $f_{B_c} = 0.489 \pm 0.005$ GeV has been obtained in lattice QCD
by the TWQCD Collaboration~\cite{Chiu:2007km}.

For $J/\psi$ meson, the explicit forms for the distribution amplitudes of twist-2
$\phi_{J/\psi}^{L,T}(x)$ and twist-3 $\phi_{J/\psi}^{t,v}(x)$ could be read as~\cite{Bondar:2004sv},
\beq
\phi_{J/\psi}^{L}(x) &=& \phi_{J/\psi}^{T}(x)=9.58\frac{f_{J/\psi}}{2\sqrt{2N_{c}}}x(1-x)
\biggl[\frac{x(1-x)}{1-2.8x(1-x)}\biggl]^{0.7}\;,
\label{eq:das-psi-LT}
\non
\phi_{J/\psi}^{t}(x) &=& 10.94\frac{f_{J/\psi}}{2\sqrt{2N_{c}}}(1-2x)^{2}
\biggl[\frac{x(1-x)}{1-2.8x(1-x)}\biggl]^{0.7}\;,
\label{eq:da-psi-t}
\\
\phi_{J/\psi}^{v}(x) &=& 1.67\frac{f_{J/\psi}}{2\sqrt{2N_{c}}}[1+(2x-1)^{2}]
\biggl[\frac{x(1-x)}{1-2.8x(1-x)}\biggl]^{0.7}\;,
\label{eq:da-psi-v}
\nonumber
\eeq
where $f_{J/\psi}$ is the decay constant with value $0.405 \pm 0.014$ GeV.

For light mesons, the distribution amplitudes for pseudoscalars, vectors, axial-vectors,
scalars, and tensors have been obtained in the QCD sum rules~\cite{Cheng:2010hn}.

\begin{itemize}
\item { for light pseudoscalars ($P$): pion and kaon }

\hspace{0.2cm}
The light-cone distribution amplitudes $\phi_P^A$ (twist-2), and $\phi_P^P$ and
$\phi_P^T$ (twist-3) have been parametrized as~\cite{Chernyak:1983ej,Ball:1998tj,Braun:2004vf}
\beq
\phi_{P}^A(x) &=& \frac{f_{P}}{2\sqrt{2N_c}}\, 6x(1-x) \left[1
+ a_1^{P} C_1^{3/2}(2x-1)
+ a_2^{P}C_2^{3/2}(2x-1)+a_4^{P}C_4^{3/2}(2x-1)\right] \;,
\label{eq:pionda-A}
\non
\phi^P_{P}(x) &=& \frac{f_{P}}{2\sqrt{2N_c}}\, \bigg[ 1
+\left(30\eta_3 -\frac{5}{2}\rho_{P}^2\right) C_2^{1/2}(2x-1)
\non &
& \hspace{35mm} -\, 3\left( \eta_3\omega_3 +
\frac{9}{20}\rho_{\pi}^2(1+6a_2^{P}) \right) C_4^{1/2}(2x-1)
\bigg]\;,
\label{eq:pionda-P}
\eeq
\beq
\phi^T_{P}(x) &=& \frac{f_{P}}{2\sqrt{2N_c}}\,
(1-2x)\bigg[ 1 + 6\left(5\eta_3 -\frac{1}{2}\eta_3\omega_3 -
\frac{7}{20}
      \rho_{P}^2 - \frac{3}{5}\rho_{P}^2 a_2^{P} \right)
(1-10x+10x^2) \bigg]\;, \nonumber
\label{eq:pionda-T}
\eeq
with the decay constants $f_\pi = 0.131$~GeV and $f_K = 0.16$~GeV; the Gegenbauer moments
$a_1^{\pi}=0, a_1^{K}=0.17 \pm 0.17, a_2^{P}= 0.115 \pm 0.115, a_4^{P}=-0.015$; the mass ratio
$\rho_{\pi(K)}=m_{\pi(K)}/m_{0}^{\pi(K)}$; $m_0^\pi=1.4$ GeV and $m_0^K = 1.6$ GeV being
the chiral masses; and the Gegenbauer polynomials $C_n^{\nu}(t)$,
\begin{eqnarray}
C_1^{3/2}(t)\; &=&\; 3\; t\;, \non
C_2^{1/2}(t)\,& =&\, \frac{1}{2} \left(3\, t^2-1\right) ,\;\; \qquad
C_2^{3/2}(t)\, =\, \frac{3}{2} \left(5\, t^2-1\right) , \;\;
\\
C_4^{1/2}(t)\, &=&\, \frac{1}{8} \left(3-30\, t^2+35\, t^4\right) , \qquad C_4^{3/2}(t) \,=\,
\frac{15}{8} \left(1-14\, t^2+21\, t^4\right) .\nonumber
\end{eqnarray}
In the above distribution amplitudes for kaon, the momentum fraction $x$ is carried by the
$s$ quark (This definition is same for the strange mesons in the following items).
We choose the parameters $\eta_3=0.015$ and
$\omega_3=-3$~\cite{Chernyak:1983ej,Ball:1998tj} for both pion and kaon, .

\item { for light vectors ($V$): $\rho$ and $K^*$}

\hspace{0.2cm}
The twist-2 distribution amplitudes for the longitudinally and transversely polarized vector
meson can be parametrized as~\cite{Ball:vector}:
\beq
\phi_{V}(x)&=&\frac{3f_{V}}{\sqrt{2N_c}} x
(1-x)\left[1+a_{1V}^{||}\, C_1^{3/2}(2x-1)+ a_{2V}^{||}\, C_2^{3/2}(2x -1)\right]\;,\label{eq:ldav}\\
\phi_{V}^T(x)&=&\frac{3f^T_{V}}{\sqrt{2N_c}} x
(1-x)\left[1+a_{1V}^{\perp}\, C_1^{3/2}(2x-1)+ a_{2V}^{\perp}\,
 C_2^{3/2}(2x -1)\right]\;,\label{eq:tdav}
\eeq
Here $f_{V}$ and $f_V^T$ are the decay constants of the vector meson in longitudinal and
transverse polarizations, respectively.
The decay constants and Gegenbauer moments for light vectors at scale $\mu=1$~GeV have been studied extensively
in the literature~\cite{Ball:2007rt,Ball:vector} and could be
found in the following Table~\ref{tab:vectors}. The masses of $\rho$ and $K^*$ are taken as 0.775~GeV and
0.892~GeV, respectively.
%%================================================
\begin{table}[hbt]
\caption{ Decay constants(in GeV) and Gegenbauer moments for light vectors.}
\label{tab:vectors}
\begin{center}\vspace{-0.3cm}{\footnotesize
\begin{tabular}[t]{c|c|c|c|c|c}
\hline\hline
$f_\rho$ & $f_\rho^T$ & $a_{1\rho}^{\parallel}$ &$a_{2\rho}^{\parallel}$
&$a_{1\rho}^\perp$ & $a_{2\rho}^\perp$   \\
\hline
$0.107\pm 0.006$&$0.105\pm 0.021$ & $\cdots$ & $0.15 \pm 0.07$ & $\cdots$ & $0.14 \pm 0.06$\\
\hline
$f_{K^*}$  &$f_{K^*}^T$ & $a_{1K^*}^\parallel$ & $a_{2K^*}^\parallel$
& $a_{1K^*}^\perp$ & $a_{2K^*}^\perp$\\
\hline
$0.118\pm 0.005$ &$0.077\pm 0.014$ & $0.03 \pm 0.02$ & $0.11 \pm 0.09$
&$0.04 \pm 0.03$ &$0.10 \pm 0.08$ \\
\hline \hline
\end{tabular}}
\end{center}
\end{table}
%%=========================================================

\hspace{0.2cm}
The asymptotic forms of the twist-3 distribution amplitudes
$\phi^{t,s}_V$ and $\phi_V^{v,a}$ are \cite{Li05:kphi}: \beq
\phi^t_V(x) &=& \frac{3f^T_V}{2\sqrt {2N_c}}(2x-1)^2,\;\;\;\;\;\;\;\;\;\;\;
  \hspace*{0.5cm} \phi^s_V(x)=-\frac{3f_V^T}{2\sqrt {2N_c}} (2x-1)~,\\
\phi_V^v(x)&=&\frac{3f_V}{8\sqrt{2N_c}}(1+(2x-1)^2),\;\;\; \ \ \
 \phi_V^a(x)=-\frac{3f_V}{4\sqrt{2N_c}}(2x-1).
\eeq

\item{ for light scalars ($S$): $a_0$, $\kappa$,
$a_0(1450)$ and $K_0^*(1430)$}

\hspace{0.2cm}
In general, the leading-twist light-cone distribution amplitude $\phi_{S}(x,\mu)$ can be expanded
as the Gegenbauer polynomials~\cite{Cheng:2005nb,Li:2008tk}:
\beq
\phi_{S}(x,\mu)&=&\frac{3}{\sqrt{2N_c}}x(1-x)\biggl\{f_{S}(\mu)+\bar f_{S}(\mu)\sum_{m=1}^\infty
B_m(\mu)C^{3/2}_m(2x-1)\biggr\},
\eeq
where $f_S(\mu)$ and $\bar f_S(\mu)$, $B_m(\mu)$, and $C_m^{3/2}(t)$ are the vector and scalar
decay constants, Gegenbauer moments, and Gegenbauer polynomials for the scalars, respectively.

\hspace{0.2cm}
For charged scalar mesons, there exists a relation between the vector and the scalar decay constants,
\beq
 \bar f_S &=& \mu_S f_S \;\;\;\;  {\rm and} \;\;\;\;  \mu_S = \frac{m_S}{m_2(\mu)-m_1(\mu)}\;,
 \label{eq:vd-sd}
\eeq
where $m_1$ and $m_2$ are the running current quark masses in the related scalars.

\hspace{0.2cm}
The values for scalar decay constants and Gegenbauer moments in the scalar meson
distribution amplitudes have been investigated at scale $\mu=1$~GeV in Ref.~\cite{Cheng:2005nb}
and are collected in Table~\ref{tab:scalars}. The masses of the considered scalars are $m_{a_0} = 0.98$~GeV,
$m_{\kappa} = 0.845$~GeV, $m_{a_0(1450)} = 1.474$~GeV, and $m_{K_0^*(1430)} = 1.425$~GeV, respectively.
%%================================================
\begin{table}[hbt]
\caption{ Scalar decay constant $\bar f_S$ (in~GeV)
and Gegenbauer moments $B_{1,3}$ for light scalars
}
\label{tab:scalars}
\begin{center}\vspace{-0.3cm}{\footnotesize
\begin{tabular}[t]{c|c|c|c}
\hline\hline
Scalars &  $\bar f_S$ & $B_1$
&  $B_3$ \\
\hline
$a_0$ &$\hspace{2.0mm}0.365 \pm 0.020$
& $-0.93 \pm 0.10$ & $\hspace{2mm}0.14 \pm 0.08$ \\
 \hline
$\kappa$ & $\hspace{2mm}0.340 \pm 0.020$
&$-0.92 \pm 0.11$ &$\hspace{2mm}0.15 \pm 0.09$\\
 \hline
$a_0(1450)$
     &$\begin{array}{c} -0.280 \pm 0.030 \;
     %\;\; ({\rm S_1})
 \\   \hspace{2.8mm}0.460 \pm 0.050 \;
 %\;\; ({\rm S_2})
       \end{array}$
     &$\begin{array}{c} \hspace{2.8mm}0.89 \pm 0.20 \;
 \\  -0.58 \pm 0.12
       \end{array}$
     &$\begin{array}{c} -1.38 \pm 0.18 \;
 \\  -0.49 \pm 0.15
       \end{array}$ \\
 \hline
$K_0^*(1430)$
     &$\begin{array}{c} -0.300 \pm 0.030 \;
     %\;\; ({\rm S_1})
 \\  \hspace{2.8mm} 0.445 \pm 0.050 \;
 %\;\; ({\rm S_2})
       \end{array}$
     &$\begin{array}{c} \hspace{2.8mm}0.58 \pm 0.07 \;
 \\  -0.57 \pm 0.13
       \end{array}$
     &$\begin{array}{c} -1.20 \pm 0.08 \;
 \\  -0.42 \pm 0.22
       \end{array}$ \\
\hline \hline
\end{tabular}}
\end{center}
\end{table}
%%=========================================================

\hspace{0.2cm}
As for the twist-3 distribution amplitudes $\phi_{S}^S$ and $\phi_{S}^T$, we adopt the following
asymptotic forms:
\beq
\phi^S_{S}&=& \frac{1}{2\sqrt {2N_c}}\bar f_{S}\;,\qquad
\phi_{S}^T=
\frac{1}{2\sqrt {2N_c}}\bar f_{S}(1-2x).
\eeq
Notice that, as inferred from Ref.~\cite{Chen:2021dwn}, the
Gegenbauer polynomials of twist-3 distribution amplitudes for light scalars are only available
in S2~\cite{Lu:2006fr} and
could mainly modify {\it CP} asymmetries in the $B$-meson decays. Because of no {\it CP} violations in the considered
$B_c$-meson decays, we will left this issue for future investigations.

\item{ for light axial-vectors ($A$): $a_1$, $b_1$, $K_{1A}$ and $K_{1B}$}

\hspace{0.2cm}
More discussions on light-cone distribution amplitudes of the light axial-vectors have
been made in the literature~\cite{Yang:2007zt,Li:2009tx}. Here, we just
simply collect the expressions adopted in this work. The details about these distribution
amplitudes could be found, e.g., in Ref.~\cite{Yang:2007zt}. The twist-2 distribution
amplitudes for the longitudinally and transversely polarized axial-vector $1^3\!P_1$
and $1^1\!P_1$ mesons can be parametrized as~\cite{Li:2009tx},
\beq
 \phi_A(x) & = & \frac{3 f}{ \sqrt{2 N_c}}  x (1- x) \left[ a_{0A}^\parallel +
a_{1A}^\parallel\, C_1^{3/2}(2x-1) +
a_{2A}^\parallel\, C_2^{3/2}(2x-1) \right] ,\label{eq:ldaa}
\\
 \phi_A^T(x) & = & \frac{3 f}{ \sqrt{2 N_c}}  x (1- x) \left[ a_{0A}^\perp +  a_{1A}^\perp\,  C_1^{3/2}(2x-1) +
a_{2A}^\perp\, C_2^{3/2}(2x-1) \right], \label{eq:tdaa}
\eeq
As for twist-3 distribution amplitudes for axial-vector meson, we use the following form~\cite{Li:2009tx}:
\beq
\phi_{A}^t(x) &= &\frac{3 f}{2\sqrt{2N_c}}\left\{ a_{0A}^\perp (2x-1)^2+ \frac{1}{2}\,a_{1A}^\perp\,(2x-1) (3 (2x-1)^2-1) \right\}
 ,\\
\phi_{A}^s(x)&=& \frac{3 f}{2\sqrt{2N_c}} \frac{d}{dx}\left\{ x (1- x) ( a_{0A}^\perp + a_{1A}^\perp (2x-1) ) \right\}.
\\
\phi_{A}^v(x)&=&\frac{3 f}{4\sqrt{2N_c}} \left\{ \frac{1}{2} a_{0A}^\parallel (1+(2x-1)^2) +  a_{1A}^\parallel (2x-1)^3 \right\}
 , \\
\phi_{A}^a(x)&=& \frac{3 f}{4\sqrt{2N_c}}\frac{d}{dx}  \left\{ x (1- x) ( a_{0A}^\parallel + a_{1A}^\parallel (2x-1))  \right\}\;.
\eeq
where $f$ is the ``normalization" decay constant (More related discussions could be found in~\cite{Li:2009tx}).

\hspace{0.2cm}
The decay constants and Gegenbauer moments have been studied extensively in the literature (see, e.g.,
Ref.~\cite{Yang:2007zt} and references therein), here we adopt the values at scale $\mu=1$~GeV as collected
in Table~\ref{tab:a-vectors}. Moreover, the masses of related axial-vectors are $m_{a_1} = 1.23$~GeV,
$m_{b_1} = 1.23$~GeV, $m_{K_{1A}} = 1.32$~GeV, and $m_{K_{1B}} = 1.34$~GeV, respectively.
%%================================================
\begin{table}[hbt]
\caption{ Decay constants (in GeV) and Gegenbauer moments for light
axial-vectors.}
\label{tab:a-vectors}
\begin{center}\vspace{-0.3cm}{\footnotesize
\begin{tabular}[t]{c|c|c|c|c|c|c}
\hline\hline
$f_{a_1}$ & $a_{0a_1}^\parallel$ & $a_{1a_1}^{\parallel}$ &$a_{2a_1}^{\parallel}$ &$a_{0a_1}^\perp$
&$a_{1a_1}^\perp$ & $a_{2a_1}^\perp$   \\
\hline
$0.238 \pm 0.010$ & $1$ &$\cdots$ & $-0.02 \pm 0.02$ & $0$ & $-1.04 \pm 0.34$ & $\cdots$
\\ \hline
$f_{b_1}$ & $a_{0b_1}^\parallel$ & $a_{1b_1}^{\parallel}$ &$a_{2b_1}^{\parallel}$ &$a_{0b1}^\perp$
&$a_{1b_1}^\perp$ & $a_{2b_1}^\perp$   \\
\hline
$0.180\pm 0.008$&$0.0028\pm 0.0026$ &$-1.95\pm 0.35$ & $\cdots$ & $1$ & $\cdots$ & $0.03 \pm 0.19$
\\ \hline
$f_{K_{1A}}$ & $a_{0K_{1A}}^\parallel$ & $a_{1K_{1A}}^{\parallel}$ &$a_{2K_{1A}}^{\parallel}$ &$a_{0K_{1A}}^\perp$
&$a_{1K_{1A}}^\perp$ & $a_{2K_{1A}}^\perp$   \\
\hline
$0.250\pm 0.013$ & $1$ & $0.00\pm 0.26$ & $-0.05 \pm 0.03$ & $0.08 \pm 0.09$
& $-1.08 \pm 0.48$ & $0.02 \pm 0.20$
\\ \hline
 $f_{K_{1B}}$  &$a_{0K_{1B}}^\parallel$ & $a_{1K_{1B}}^\parallel$ & $a_{2K_{1B}}^\parallel$
 &$a_{0K_{1B}}^\perp$
& $a_{1K_{1B}}^\perp$ & $a_{2K_{1B}}^\perp$\\
\hline
$0.190\pm 0.010$ &$0.14 \pm 0.15$ & $-1.95 \pm 0.45$ & $0.02 \pm 0.10$
&$1$ &$0.17 \pm 0.22$ & $-0.02 \pm 0.22$ \\
\hline \hline
\end{tabular}}
\end{center}
\end{table}
%%=========================================================

\item { for light tensors ($T$): $a_2$ and $K_2^*$}

\hspace{0.2cm}
Here, we present the light-cone
distribution amplitudes of light tensor mesons following Ref.~\cite{Cheng:2010hn,Wang:2010ni}:
\beq
\phi_{T}(x)&=&\frac{3f_{T}}{\sqrt{2N_c}} \phi_\parallel(x)\;,\qquad
\phi_{T}^T(x)=\frac{3f^T_{T}}{\sqrt{2N_c}} \phi_\perp(x)\;,\non
\phi_{T}^t(x)&=&\frac{f_{T}^T}{2\sqrt{2N_c}} h_\parallel^t(x)\;,\qquad
\phi_{T}^s(x)=\frac{f_{T}^T}{4\sqrt{2N_c}} \frac{d}{dx}h_\parallel^s(x)\;,\\
\phi_{T}^v(x)&=&\frac{f_{T}}{2\sqrt{2N_c}} g_\parallel^v(x)\;,\qquad
\phi_{T}^a(x)=\frac{f_{T}}{8\sqrt{2N_c}} \frac{d}{dx}g_\perp^a(x)\;, \nonumber
\eeq
with
\beq
\phi_\parallel(x)&=&\phi_\perp(x)=x(1-x)[a_1\;C_1^{3/2}(t)]\;,
\label{eq:twist-2} \non
h_\parallel^t(x)&=& \frac{15}{2}(1-6x+6x^2)t, \qquad
h_\parallel^s(x)= 15x(1-x)t\;,\\
g_\perp^v(x)&=&5t^3, \qquad g_\perp^a(x)=20x(1-x)t\;. \nonumber
\eeq
with the Gegenbauer moment $a_1=\frac{5}{3}$ for the first rough estimates.
It is worth commenting that, in principle, the Gegenbauer moments for $a_2$ and $K_2^*$
should usually be different due to the expected SU(3)-flavor symmetry breaking effects.
Therefore, the larger Gegenbauer moment $a_1$ adopted here
will demand further improvements through precise measurements.
The decay constants for $a_2$ and $K_2^*$ are presented in Table~\ref{tab:tensors}.
Moreover, the masses for $a_2$ and $K_2^*$ are adopted as 1.318~GeV and 1.427~GeV, respectively.

%%================================================
\begin{table}[hbt]
\caption{ Decay constants(in GeV) for light tensors.}
\label{tab:tensors}
\begin{center}\vspace{-0.3cm}{\footnotesize
\begin{tabular}[t]{cccc}
\hline\hline
$f_{a_2}$ & $f_{a_2}^T$ & $f_{K_2^*}$  &$f_{K_2^*}^T$  \\
$0.107\pm 0.006$&$0.105\pm 0.021$
&$0.118\pm 0.005$&$0.077\pm 0.014$\\
\hline \hline
\end{tabular}}
\end{center}
\end{table}
%%=========================================================
\end{itemize}

\section{\boldmath Factorization formulas for $B_c^+ \to J/\psi M^+$ decays}
\label{app:facformulas}

In this section, we present the factorization formulas explicitly for the $B_c^+ \to J/\psi M^+$
decays calculated in the iPQCD formalism. First of all, the expressions for $B_c^+ \to J/\psi
P^+$ decays could be referred to Ref.~\cite{Liu:2018kuo} for details, and are no longer
presented here. For $B_c^+ \to J/\psi S^+$ decays, the factorization formulas are presented as follows,
\begin{itemize}
    \item {For factorizable emission diagrams, }
\beq
F_e (S) &=& -8 \pi C_F m_{B_c}^4 \int_0^1 dx_1 dx_3
\int_0^\infty b_1db_1 b_3db_3 \phi_{B_c}(x_1,b_1) (r_3^2 -1)
\non && \times
\biggl\{  \left[ r_3 (r_b +2 x_3 -2) \phi_{J/\psi}^t(x_3)
-
(2 r_b + x_3 -1) \phi_{J/\psi}^L(x_3)  \right] h_a(x_1,x_3,b_1,b_3) E_f(t_a)
 \non &&
+
\left[r^2_3 (x_1 -1)-r_c \right]\phi_{J/\psi}^L(x_3) h_b(x_1,x_3,b_1,b_3)
E_f(t_b)  \biggr\} \;,
\label{eq:fe-S}
\eeq
where the ratios $r_b= m_b/m_{B_c}$ and $r_c = m_c/ m_{B_c}$.
The hard function $h_i(x_i,
b_i)$ and the evolution function $E_f(t_i)$ could refer to those expressions in
Ref.~\cite{Liu:2018kuo} and are collected in Appendix~\ref{app:rel-fun}.

    \item {For non-factorizable emission diagrams,}
\beq
M_e (S) &=& \frac{32}{\sqrt{6}}\pi C_F m_{B_c}^4 \int_0^1 dx_1 dx_2 dx_3
\int_0^\infty b_1db_1 b_2db_2 \phi_{B_c}(x_1,b_1) \phi_S(x_2) (r_3^2 -1)
\non & &
\times
\biggl\{\left[(r_3^2-1)(x_1+x_2-1)\phi_{J/\psi}^L(x_3)
+r_3 (x_3 -x_1)\phi_{J/\psi}^t(x_3)\right] E_f(t_c)
\non &&
\times
h_c(x_1,x_2,x_3,b_1,b_2)+\left[(2x_1-(x_2+x_3)+r_3^2(x_2-x_3))\phi_{J/\psi}^L(x_3)
\right.
\non &&
\left.
+ r_3 (x_3-x_1)\phi_{J/\psi}^t(x_3)\right]
h_d(x_1,x_2,x_3,b_1,b_2)E_f(t_d)\biggr\}\;.
\label{eq:nfe-S}
 \eeq

\end{itemize}

Then, for $B_c^+ \to J/\psi V^+$ decays, the factorization formulas with polarization
contributions are collected as the following,
\begin{itemize}
    \item {For factorizatable emission diagrams, }
\beq
F_e^L (V) & = & 8 \pi C_F m_{B_c}^4 \int_0^1 dx_1 dx_3
\int_0^\infty b_1db_1 b_3db_3 \phi_{B_c}(x_1,b_1) \sqrt{1 -r_3^2}
\non && \times
\biggl\{  \left[ r_3 (r_b +2 x_3 -2) \phi_{J/\psi}^t(x_3)
-
(2 r_b + x_3 -1) \phi_{J/\psi}^L(x_3)  \right] h_a(x_1,x_3,b_1,b_3)
 \non &&
\times  E_f(t_a) +
\left[r^2_3 (x_1 -1)-r_c \right]\phi_{J/\psi}^L(x_3) h_b(x_1,x_2,b_1,b_2)
E_f(t_b)  \biggr\} \;,
\label{eq:fe-VL}
\\
F_e^N (V) & = & 8 \pi C_F m_{B_c}^4 r_2 \int_0^1 dx_1 dx_3
\int_0^\infty b_1db_1 b_3db_3 \phi_{B_c}(x_1,b_1)
\non && \times
\biggl\{  \left[ ( r_3^2 (r_b +4 x_3 -2) +r_b -2 ) \phi_{J/\psi}^T(x_3)
- r_3 (
(4 r_b + x_3 (1+ r^2_3) -2) ) \right.
 \non &&
\left.  \times \phi_{J/\psi}^v(x_3)  \right]
h_a(x_1,x_3,b_1,b_3) E_f(t_a)
- r_3
\left[r^2_3 + 2 r_c - 2 x_1 +1 \right]\phi_{J/\psi}^v(x_3)
 \non &&
 \times
 h_b(x_1,x_2,b_1,b_2)
E_f(t_b)  \biggr\} \;,
\label{eq:fe-VN}
\eeq
\beq
F_e^T (V) & = & 16 \pi C_F m_{B_c}^4 r_2 \int_0^1 dx_1 dx_3
\int_0^\infty b_1db_1 b_3db_3 \phi_{B_c}(x_1,b_1)
\non && \times
\biggl\{  \left[ (r_b -2 ) \phi_{J/\psi}^T(x_3)
+ r_3  x_3 \phi_{J/\psi}^v(x_3)  \right]  h_a(x_1,x_3,b_1,b_3) E_f(t_a)
 \non &&
- r_3 \phi_{J/\psi}^v(x_3) h_b(x_1,x_2,b_1,b_2)
E_f(t_b)  \biggr\} \;,
\label{eq:fe-VT}
\eeq

\item {For non-factorizable emission diagrams, }
\beq
M_e^L (V) &=& \frac{32}{\sqrt{6}}\pi C_F m_{B_c}^4 \int_0^1 dx_1 dx_2 dx_3
\int_0^\infty b_1db_1 b_2db_2 \phi_{B_c}(x_1,b_1) \phi_V(x_2) \sqrt{1-r_3^2}  \non & & \times
\biggl\{\left[(r_3^2-1)(x_1+x_2-1)\phi_{J/\psi}^L(x_3)
+r_3 (x_3 -x_1)\phi_{J/\psi}^t(x_3)\right] E_f(t_c)
\non &&
\times  h_c(x_1,x_2,x_3,b_1,b_2)
+\left[(2x_1-(x_2+x_3)+r_3^2(x_2-x_3))\phi_{J/\psi}^L(x_3)
\right.
\non &&
\left.
+ r_3 (x_3-x_1)\phi_{J/\psi}^t(x_3)\right] h_d(x_1,x_2,x_3,b_1,b_2)E_f(t_d)\biggr\}\;,
\label{eq:nfe-VL}
\eeq
\beq
M_e^N (V) &=& -\frac{32}{\sqrt{6}}\pi C_F m_{B_c}^4 \int_0^1 dx_1 dx_2 dx_3
\int_0^\infty b_1db_1 b_2db_2 \phi_{B_c}(x_1,b_1) r_2
\non & & \times
\biggl\{\left[(r_3^2(x_1 -x_2 -2 x_3 +1)+ x_1 + x_2 -1 )\phi_V^v(x_2)
+ (1- r_3^2)^2 (x_1 +x_2 -1) \right.
\non &&
\left.
\times
\phi_V^a(x_2)\right] \phi_{J/\psi}^T(x_3) E_f(t_c)
h_c(x_1,x_2,x_3,b_1,b_2)
+\left[( (r_3^2(x_1+x_2 - 2 x_3) + x_1
\right.
\non &&
\left.
- x_2) \phi_V^v(x_2)+ (1-r_3^2)^2 (x_1 -x_2) \phi_V^a(x_2))\phi_{J/\psi}^T(x_3)
+ 2 r_3 ((x_3-x_2) r_3^2
\right.
\non &&
\left.
 +x_2 +x_3 -2 x_1 )\phi_V^v(x_2)\phi_{J/\psi}^v(x_3)\right] h_d(x_1,x_2,x_3,b_1,b_2)E_f(t_d)\biggr\}\;,
\label{eq:nfe-VN}
\\
M_e^T (V) &=& -\frac{64}{\sqrt{6}}\pi C_F m_{B_c}^4 \int_0^1 dx_1 dx_2 dx_3
\int_0^\infty b_1db_1 b_2db_2 \phi_{B_c}(x_1,b_1) r_2
\non & & \times
\biggl\{\left[(r_3^2(x_1 -x_2 -2 x_3 +1)+ x_1 + x_2 -1 )\phi_V^a(x_2)
+  (x_1 +x_2 -1) \right.
\non &&
\left.
\times
\phi_V^v(x_2)\right] \phi_{J/\psi}^T(x_3) E_f(t_c)
h_c(x_1,x_2,x_3,b_1,b_2)
+\left[ ( ( x_1 - x_2) \phi_V^v(x_2)
\right.
\non &&
\left.
 + (  x_1 (1 +r_3^2)^2 + x_2 (r_3^2 -1) - 2 r_3^2 x_3) \phi_V^a(x_2)) \phi_{J/\psi}^T(x_3)
+ 2 r_3 ( x_3 (r_3^2 +1 )
\right.
\non &&
\left.
 -x_2 (r_3^2 -1) -2 x_1 )\phi_V^v(x_2)\phi_{J/\psi}^v(x_3)\right]
 h_d(x_1,x_2,x_3,b_1,b_2)E_f(t_d)\biggr\}\;.
\label{eq:nfe-VT}
\eeq
\end{itemize}

For $B_c^+ \to J/\psi A^+$ decays, the factorization formulas could be easily obtained as
\beq
F_A^h &=& - F_V^h\;,  \qquad   M_A^h = - M_V^h \;,
\label{eq:A}
\eeq
associated with the replacements of $\phi_V \to \phi_A$ and $r_V \to r_A$ correspondingly.

And, for $B_c^+ \to J/\psi T^+$ decays, as discussed in Ref.~\cite{Liu:2017cwl}, the related
factorization formulas could be straightforwardly obtained through those for $B_c^+ \to J/\psi
V^+$ decays as follows,
\beq
M_T^L &=&  \sqrt{\frac{2}{3}} \; M_V^L\;, \qquad
M_T^{N,T} = \sqrt{\frac{1}{2}} \; M_V^{N, T}\;.
\label{eq:T}
\eeq
in association with the corresponding replacements of $\phi_V \to \phi_T$ and $r_V \to r_T$.

\section{\boldmath  Related functions}
\label{app:rel-fun}
We here collect the related functions, i.e., hard functions $h_i(x_i, b_i)$ and evolution
functions $E_f(t_i)$, in the factorization formulas.

The general form of $h_i(x_i, b_i)$ in the factorization formulas could be written as follows,
\beq
h_{a, b}(x_1,x_3,b_1,b_3) &=& \left[
\theta(b_3-b_1)I_0(\sqrt{\beta_{a,b}}b_1) K_0(\sqrt{\beta_{a,b}}b_3)
+(b_1 \leftrightarrow b_3)\right]K_0(\sqrt{\alpha}b_1), \\
 h_{c,d}(x_1,x_2,x_3,b_1,b_2) &=&
 \left[\theta(b_2-b_1) I_0(\sqrt{\alpha} b_1)
 K_0(\sqrt{\alpha} b_2)
+(b_1 \leftrightarrow b_2) \right]
K_0( \sqrt{\beta_{c,d}} b_2),
\label{eq:pp1}
 \eeq
with the factors $\alpha$ and $\beta_{a,b,c,d}$ and the hard scales $t_{a,b,c,d}$,
\beq
\alpha &=& -[(x_1-x_3 (1-r_2^2))(x_1-x_3 r_3^2)]m_{B_c}^2,\\
\beta_a &=& -[(1-x_3 (1- r_2^2))(1-x_3 r_3^2)-r_b^2]m_{B_c}^2,\qquad
\beta_b = -[(1-x_1-r_2^2)(r_3^2-x_1)-r_c^2]m_{B_c}^2,\\
\beta_c &=& -[(x_3 r_3^2+(1-x_2)(1-r_3^2)-x_1)
(x_3(1-r_2^2)+(1-x_2) r_2^2-x_1)]m_{B_c}^2 , \\
\beta_d &=& -[(x_2 r_2^2+x_3 (1-r_2^2)-x_1)
(x_3 r_3^2 + x_2 (1-r_3^2)-x_1)]m_{B_c}^2,\\
t_a &=& \max(\sqrt{|\alpha|},\sqrt{|\beta_a|},1/b_1,1/b_3),
\qquad
t_b = \max(\sqrt{|\alpha|},\sqrt{|\beta_b|},1/b_1,1/b_3),\\
t_c &=& \max(\sqrt{|\alpha|}, \sqrt{|\beta_c|}, 1/b_1, 1/b_2) ,
\qquad
t_d = \max(\sqrt{|\alpha|}, \sqrt{|\beta_d|}, 1/b_1, 1/b_2).
\eeq
Note that, as $\alpha$ and $\beta_{a,b,c,d}$ are negative, the associated
Bessel functions transform as
\begin{eqnarray}
K_0(\sqrt{y}) = K_0(-i\sqrt{|y|})= \frac{i \pi}{2} [J_0(\sqrt{|y|})
+ i N_0(\sqrt{|y|})] \;, \qquad I_0(\sqrt{y}) = J_0(\sqrt{|y|}),  \hspace{0.5cm}
\end{eqnarray}
for $y<0$.

The evolution functions $E_f(t)\equiv \alpha_s(t)C_i(t) S_i(t) $ contain
the Wilson coefficients
\beq
  C_{ab}(t) &=&  \frac{1}{3} C_1(t) + C_2(t) \;,
  \qquad
  C_{cd}(t) = C_1(t)\;,
\eeq
and the Sudakov factors
\beq
S_{ab}(t)
&=& s_c\left(x_1 P_1^+, b_1\right) +s_c\left(x_3
P_3^-, b_3\right) +s_c\left((1-x_3) P_3^-,
b_3\right) \non
&& -\frac{1}{\beta_1}\left[\frac{11}{6} \ln\frac{\ln(t/\Lambda)}{\ln(m_c/\Lambda)}
\right],
\label{eq:sab}\\
S_{cd}(t) &=& s_c\left(x_1 P_1^+, b_1\right)
 +s\left(x_2 P_2^+, b_2\right)
+s\left((1-x_2) P_2^+, b_2\right) +s_c\left(x_3
P_3^-, b_1\right) \non
 && +s_c\left((1-x_3) P_3^-,
b_1\right) -\frac{1}{\beta_1}\left[
\frac{11}{6} \ln\frac{\ln(t/\Lambda)}{\ln(m_c/\Lambda)}
+\ln\frac{\ln(t/\Lambda)}{-\ln(b_2\Lambda)}\right]\;,
\label{eq:scd}
\eeq
where the explicit expression of the Sudakov exponent $s(Q,b)$ for an energetic light
quark is referred to Ref.~\cite{Keum:2000ph} and that of the other Sudakov
factor $s_c(Q,b)$ with inclusion of finite charm quark mass effects can be found in
Appendix~\ref{app:SF-bc}.

\section{\boldmath  Sudakov factor $S_c(Q,b)$ for $B_c$-meson decays}
\label{app:SF-bc}

Here, we show the explicit form of newly derived Sudakov factor at next-to-leading
logarithm accuracy with two-loop running coupling constant $\alpha_s$ that could
make the framework of iPQCD formalism for $B_c$-meson decays more self-consistent.
\beq
S_c(Q,b) &=& \frac{a_1}{2 \beta_1} \biggl\{ \hat{Q} \ln\hat{Q} -
\hat{c}\ln{\hat{c}} -(\hat{Q}-\hat{c})(1+\ln{\hat{b}})\biggr\} \non
 && +\frac{a_2}{4 \beta_1^2} \biggl\{-\ln{\frac{\hat{Q}}{\hat{c}}} + \frac{\hat{Q}-\hat{c}}{\hat{b}}\biggr\}
+ \frac{a_1}{4 \beta_1} {\ln{\frac{\hat{Q}}{\hat{c}}}} \ln{\frac{e^{2\gamma_E -1}}{2}}\non
&& +\frac{a_1\beta_2}{4\beta_1^3}\biggl\{\ln\frac{\hat{Q}}{\hat{c}}+\frac{1}{2}
(\ln^22\hat{Q} - \ln^22\hat{c})-\frac{1}{\hat{b}}(1+\ln2\hat{b})(\hat{Q}-\hat{c})\biggr\}\non
&&- \frac{a_2\beta_2}{4\beta_1^4}\biggl\{\frac{1}{4\hat{b}^2}(1+2\ln2\hat{b})(\hat{Q}-\hat{c})
+\frac{3}{4}(\frac{1}{\hat{Q}}-\frac{1}{\hat{c}})+\frac{1}{2}(\frac{\ln2\hat{Q}}{\hat{Q}}
-\frac{\ln2\hat{c}}{\hat{c}})\biggr\}
\non
&&+\frac{a_2\beta_2^2}{16\beta_1^6}\biggl\{(\frac{2}{27}
+\frac{2}{9}\ln2\hat{b}
+\frac{1}{3}\ln^22\hat{b})
\frac{\hat{Q}-\hat{c}}{\hat{b}}
+\frac{19}{108}(\frac{1}{\hat{Q}^2}-\frac{1}{\hat{c}^2})
+\frac{5}{18}(\frac{\ln2\hat{Q}}{\hat{Q}^2}-\frac{\ln2\hat{c}}{\hat{c}^2})\non
&&
+\frac{1}{6}(\frac{\ln^22\hat{Q}}{\hat{Q}^2}-\frac{\ln^22\hat{c}}{\hat{c}^2})\biggr\}
+\frac{a_1\beta_2}{8\beta_1^3}\biggl\{\frac{1}{\hat{Q}}
-\frac{1}{\hat{c}}+\frac{\ln{2\hat{Q}}}{\hat{Q}}
-\frac{\ln2\hat{c}}{\hat{c}}\biggr\}\ln\frac{e^{2\gamma_E-1}}{2}\;,
\label{eq:aSuda-1}
\eeq
with definitions: $\hat{Q} \equiv \ln{[x P^+/\Lambda]}$, $\hat{c} \equiv
\ln{[m_c/(\sqrt{2}\Lambda)]}$, and $\hat{b} \equiv \ln{[1/(b\Lambda)]}$, and the constants
$a_1 = C_F = \frac{4}{3}$, $a_2 = \frac{67}{9} - \frac{\pi^2}{3} - \frac{10}{27}n_f$,
$\beta_1 = \frac{33-2n_f}{12}$, and $\beta_2= \frac{153-19n_f}{24}$ with $n_f$ being the
flavor number. Notice that, when the replacement $\hat{c} \to \hat{b}$ is adopted,
then the formula presented in Eq.~(\ref{eq:aSuda-1}) will recover the Sudakov factor
for $B$-meson decays with strong running coupling constant $\alpha_s$ at two-loop level~\cite{Keum:2000ph}.
And furthermore, when the term $\beta_2$ in Eq.~(\ref{eq:aSuda-1}) is turned off,
the equation will then return to the Sudakov factor with one-loop running coupling
constant $\alpha_s$ that has been adopted in this work,
\beq
S_c(Q,b) &=& \frac{a_1}{2 \beta_1} \biggl\{ \hat{Q} \ln\hat{Q} -
\hat{c}\ln{\hat{c}} -(\hat{Q}-\hat{c})(1+\ln{\hat{b}})\biggr\} \non
 && +\frac{a_2}{4 \beta_1^2} \biggl\{-\ln{\frac{\hat{Q}}{\hat{c}}} + \frac{\hat{Q}-\hat{c}}{\hat{b}}\biggr\}
+ \frac{a_1}{4 \beta_1} {\ln{\frac{\hat{Q}}{\hat{c}}}} \ln{\frac{e^{2\gamma_E -1}}{2}}\;.
\label{eq:aSuda-2}
\eeq

\end{appendix}

%%====================================================================
%%%%%%%%%%%%%%%%%%%%    References  ==================================
%%%%==================================================================

\end{CJK*}

\begin{thebibliography}{99}

%\cite{CDF:1998axz}
\bibitem{CDF:1998axz}
F.~Abe \textit{et al.} (CDF Collaboration),
%``Observation of $B_c$ mesons in $p\bar{p}$ collisions at $\sqrt{s} = 1.8$ TeV,''
Phys. Rev. D \textbf{58}, 112004 (1998).
%doi:10.1103/PhysRevD.58.112004
%[arXiv:hep-ex/9804014 [hep-ex]].

%\cite{CDF:1998ihx}
\bibitem{CDF:1998ihx}
F.~Abe \textit{et al.} (CDF Collaboration),
%``Observation of the $B_c$ meson in $p\bar{p}$ collisions at $\sqrt{s} = 1.8$ TeV,''
Phys. Rev. Lett. \textbf{81}, 2432-2437 (1998).
%doi:10.1103/PhysRevLett.81.2432
%[arXiv:hep-ex/9805034 [hep-ex]].

%\cite{CDF:2006kbk}\cite{CDF:2007umr}
\bibitem{CDF:2006kbk}
A.~Abulencia \textit{et al.} (CDF Collaboration),
%``Measurement of the B(c)+ meson lifetime using B(c)+ ---\ensuremath{>} J/psi e+ nu(e),''
Phys. Rev. Lett. \textbf{97}, 012002 (2006).
%doi:10.1103/PhysRevLett.97.012002
%[arXiv:hep-ex/0603027 [hep-ex]].

%\cite{D0:2008thm}\cite{D0:2008bqs}
\bibitem{D0:2008thm}
V.M.~Abazov \textit{et al.} (D0 Collaboration),
%``Measurement of the lifetime of the $B_c^\pm$ meson in the semileptonic decay channel,''
Phys. Rev. Lett. \textbf{102}, 092001 (2009).
%doi:10.1103/PhysRevLett.102.092001
%.[arXiv:0805.2614 [hep-ex]].



%\cite{CDF:2007umr}
\bibitem{CDF:2007umr}
T.~Aaltonen \textit{et al.} (CDF Collaboration),
%``Observation of the Decay $B^+$ -($c$) $\to J/\psi \pi^\pm$ and Measurement of the $B^+$ -($c$) Mass,''
Phys. Rev. Lett. \textbf{100}, 182002 (2008).
%doi:10.1103/PhysRevLett.100.182002
%[arXiv:0712.1506 [hep-ex]].


%\cite{D0:2008bqs}
\bibitem{D0:2008bqs}
V.M.~Abazov \textit{et al.} (D0 Collaboration),
%``Observation of the $B_c$ Meson in the Exclusive Decay $B_c \to J/\psi \pi$,''
Phys. Rev. Lett. \textbf{101}, 012001 (2008).
%doi:10.1103/PhysRevLett.101.012001
%[arXiv:0802.4258 [hep-ex]].


\bibitem{Brambilla:2004wf}
  N.~Brambilla {\it et al.} (Quarkonium Working Group),
  %``Heavy quarkonium physics,''
  arXiv: hep-ph/0412158.

%\cite{Brambilla:2010cs}
\bibitem{Brambilla:2010cs}
N.~Brambilla, S.~Eidelman, B.K.~Heltsley, R.~Vogt, G.T.~Bodwin, E.~Eichten,
A.D.~Frawley, A.B.~Meyer, R.E.~Mitchell, V.~Papadimitriou \textit{et al.},
%``Heavy Quarkonium: Progress, Puzzles, and Opportunities,''
Eur. Phys. J. C \textbf{71}, 1534 (2011).
%doi:10.1140/epjc/s10052-010-1534-9
%[arXiv:1010.5827 [hep-ph]].


%\cite{Li:2018lxi}\cite{Huang:2018nnq}\cite{Cheung:2020sbq}
\bibitem{Li:2018lxi}
Y.~Li and C.D.~L\"u,
%``Recent Anomalies in B Physics,''
Sci. Bull. \textbf{63}, 267  (2018).
%doi:10.1016/j.scib.2018.02.003
%[arXiv:1808.02990 [hep-ph]].


%\cite{Huang:2018nnq}
\bibitem{Huang:2018nnq}
Z.R.~Huang, Y.~Li, C.D.~L\"u, M.A.~Paracha, and C.~Wang,
%``Footprints of New Physics in $b\to c\tau\nu$ Transitions,''
Phys. Rev. D \textbf{98}, 095018 (2018).
%doi:10.1103/PhysRevD.98.095018
%[arXiv:1808.03565 [hep-ph]].


%\cite{Cheung:2020sbq}
\bibitem{Cheung:2020sbq}
K.~Cheung, Z.R.~Huang, H.D.~Li, C.D.~L\"u, Y.N.~Mao, and R.Y.~Tang,
%``Revisit to the $b\to c\tau\nu$ transition: In and beyond the SM,''
Nucl. Phys. \textbf{B965}, 115354 (2021).
%doi:10.1016/j.nuclphysb.2021.115354
%[arXiv:2002.07272 [hep-ph]].

%\cite{Elahi:2021jia}
\bibitem{Elahi:2021jia}
F.~Elahi, G.~Elor, and R.~McGehee,
%``Charged B mesogenesis,''
Phys. Rev. D \textbf{105}, 055024 (2022).
%doi:10.1103/PhysRevD.105.055024
%[arXiv:2109.09751 [hep-ph]].


%\cite{ParticleDataGroup:2020ssz}
\bibitem{ParticleDataGroup:2020ssz}
P.A.~Zyla \textit{et al.} (Particle Data Group),
%``Review of Particle Physics,''
Prog. Theor. Eep. Phys. \textbf{2020}, 083C01 (2020), and update at https://pdglive.lbl.gov/.


%\cite{HFLAV:2022pwe}
\bibitem{HFLAV:2022pwe}
Y.~Amhis \textit{et al.} (HFLAV Collaboration),
%``Averages of $b$-hadron, $c$-hadron, and $\tau$-lepton properties as of 2021,''
%arXiv:2206.07501 [hep-ex]
Phys. Rev. D \textbf{107}, 052008 (2023); updated in
https://hflav.web.cern.ch/.



%\cite{LHCb:2012ag}
\bibitem{LHCb:2012ag}
R.~Aaij \textit{et al.} (LHCb Collaboration),
%``First observation of the decay $B_c^+ \to J/\psi \pi^+\pi^-\pi^+$,''
Phys. Rev. Lett. \textbf{108}, 251802 (2012).
%doi:10.1103/PhysRevLett.108.251802
%[arXiv:1204.0079 [hep-ex]].

%\bibitem{Khachatryan:2014nfa}
%\cite{CMS:2014oqy}
\bibitem{CMS:2014oqy}
V.~Khachatryan \textit{et al.} (CMS Collaboration),
%``Measurement of the ratio of the production cross sections times branching fractions of $B_{c}^{\pm} \to J/\psi \pi^{\pm}$ and $B^{\pm} \to J/\psi K^{\pm}$ and $\mathcal{B}(B_{c}^{\pm} \to J/\psi \pi^{\pm}\pi^{\pm}\pi^{\mp})/\mathcal{B}(B_{c}^{\pm} \to J/\psi \pi^{\pm})$ in pp collisions at $\sqrt{s} =$ 7 TeV,''
\jhep \textbf{01} (2015) 063.
%doi:10.1007/JHEP01(2015)063
%[arXiv:1410.5729 [hep-ex]].

\bibitem{Likhoded:2009ib}
  A.K.~Likhoded and A.V.~Luchinsky,
  %``Light hadron production in B(c) ---> J/psi + X decays,''
  Phys.\ Rev.\ D {\bf 81}, 014015 (2010).

\bibitem{Rakitin:2009ya}
  A.~Rakitin and S.~Koshkarev,
  %``Hadronic $B_{c}$ decays as a test of $B_{c}$ cross section,''
  Phys.\ Rev.\ D {\bf 81}, 014005 (2010).

\bibitem{Wang:2012vna}
  Z.~G.~Wang,
  %``The $B_c$-decays $B_c^+ \to J/\psi \pi^+\pi^-\pi^+$, $\eta_c \pi^+\pi^-\pi^+ $,''
  Phys.\ Rev.\ D {\bf 86}, 054010 (2012).

\bibitem{Luchinsky:2012rk}
  A.V.~Luchinsky,
  %``Production of charged $\pi$-mesons in exclusive $B_{c}\to V(P)+n\pi$ decays,''
  Phys.\ Rev.\ D {\bf 86}, 074024 (2012).

\bibitem{Qiao:2012hp}
  C.F.~Qiao, P.~Sun, D.~Yang, and R.L.~Zhu,
  %``B$_c$ exclusive decays to charmonium and a light meson at next-to-leading order accuracy,''
  Phys.\ Rev.\ D {\bf 89}, 034008 (2014).

%\cite{LHCb:2013rud}
\bibitem{LHCb:2013rud}
R.~Aaij \textit{et al.} (LHCb Collaboration),
%``Observation of the decay $B_c \rightarrow J/\psi K^+ K^- \pi^+ $,''
\jhep \textbf{11} (2013) 094.
%doi:10.1007/JHEP11(2013)094
%[arXiv:1309.0587 [hep-ex]].


%\cite{Luchinsky:2013yla}
\bibitem{Luchinsky:2013yla}
A.V.~Luchinsky,
%``Production of K mesons in exclusive $B_c$ decays,''
arXiv:1307.0953 [hep-ph].




%\cite{LHCb:2021tdf}
\bibitem{LHCb:2021tdf}
R.~Aaij \textit{et al.} (LHCb Collaboration),
%``Study of $ {\mathrm{B}}_{\mathrm{c}}^{+} $ decays to charmonia and three light hadrons,''
\jhep \textbf{01} (2022) 065;
%doi:10.1007/JHEP01(2022)065
%[arXiv:2111.03001 [hep-ex]].
%\cite{LHCb:2022ioi}
%\bibitem{LHCb:2022ioi}
%R.~Aaij \textit{et al.} [LHCb],
%``Study of $B_c^+$ meson decays to charmonia plus multihadron final states,''
%arXiv:2208.08660 [hep-ex].
 \textbf{07} (2023) 198.


%\bibitem{Aaij:2013vcx}
%\cite{LHCb:2013hwj}
\bibitem{LHCb:2013hwj}
R.~Aaij \textit{et al.} (LHCb Collaboration),
%``First observation of the decay $B_{c}^{+}\to J/\psi K^+$,''
\jhep\textbf{09} (2013) 075.
%doi:10.1007/JHEP09(2013)075
%[arXiv:1306.6723 [hep-ex]].

%\bibitem{Aaij:2016tcz}
%\cite{LHCb:2016vni}
\bibitem{LHCb:2016vni}
R.~Aaij \textit{et al.} (LHCb Collaboration),
%``Measurement of the ratio of branching fractions $\mathcal{B}(B_{c}^{+} \to J/\psi K^{+})/\mathcal{B}(B_{c}^{+} \to J/\psi\pi^{+})$,''
\jhep \textbf{09} (2016) 153.
%doi:10.1007/JHEP09(2016)153
%[arXiv:1607.06823 [hep-ex]].



%\cite{Liu:2018kuo}
\bibitem{Liu:2018kuo}
X.~Liu, H.-n.~Li, and Z.J.~Xiao,
%``Improved perturbative QCD formalism for $B_c$ meson decays,''
Phys. Rev. D \textbf{97}, 113001 (2018).
%doi:10.1103/PhysRevD.97.113001
%[arXiv:1801.06145 [hep-ph]].
%13 citations counted in INSPIRE as of 15 Jul 2022


%\cite{Liu:2020upy}
\bibitem{Liu:2020upy}
X.~Liu, H.-n.~Li, and Z.J.~Xiao,
%``Next-to-leading-logarithm $k_T$ resummation for $B_c\to J/\psi$ decays,''
Phys. Lett. B \textbf{811}, 135892 (2020).



\bibitem{Rui:2014tpa}
  Z.~Rui and Z.T.~Zou,
  %``S-wave ground state charmonium decays of $B_c$ mesons in the perturbative QCD approach,''
  Phys.\ Rev.\ D {\bf 90}, 114030 (2014).

\bibitem{Sun:2008ew}
  J.F.~Sun, D.S.~Du, and Y.L.~Yang,
  %``Study of $B_c \to J/\psi \pi$, $\eta_c \pi$ decays with perturbative QCD approach,''
  Eur.\ Phys.\ J.\ C {\bf 60}, 107 (2009).




\bibitem{Rui:2016opu}
  Z.~Rui, H.~Li, G.~x.~Wang, and Y.~Xiao,
  %``Semileptonic decays of $B_c$ meson to S-wave charmonium states in the perturbative QCD approach,''
  Eur.\ Phys.\ J.\ C {\bf 76}, 564 (2016).


%\cite{Chang:1992pt}
\bibitem{Chang:1992pt}
  C.H.~Chang and Y.Q.~Chen,
%  ``The Decays of B(c) meson,''
  Phys.\ Rev.\ D {\bf 49}, 3399 (1994).


\bibitem{Liu:1997hr}
  J.F.~Liu and K.T.~Chao,
  %``$B_c$ meson weak decays and CP violation,''
  Phys.\ Rev.\ D {\bf 56}, 4133 (1997).


\bibitem{Colangelo:1999zn}
  P.~Colangelo and F.~De Fazio,
  %``Using heavy quark spin symmetry in semileptonic $B_c$ decays,''
  Phys.\ Rev.\ D {\bf 61}, 034012 (2000).

%\cite{AbdElHady:1999xh}
\bibitem{AbdElHady:1999xh}
  A.~Abd El-Hady, J.H.~Mu\~{n}oz and J.P.~Vary,
%  ``Semileptonic and nonleptonic B(c) decays,''
  Phys.\ Rev.\ D {\bf 62}, 014019 (2000).
%  doi:10.1103/PhysRevD.62.014019
%  [hep-ph/9909406].


%\cite{Verma:2001hb}
\bibitem{Verma:2001hb}
R.C.~Verma and A.~Sharma,
%``Quark diagram analysis of weak hadronic decays of B/c+ meson,''
Phys. Rev. D \textbf{65}, 114007 (2002).



%\cite{Ebert:2003cn}
\bibitem{Ebert:2003cn}
  D.~Ebert, R.N.~Faustov and V.O.~Galkin,
%  ``Weak decays of the $B_c$ meson to charmonium and $D$ mesons
%    in the relativistic quark model,''
  Phys.\ Rev.\ D {\bf 68}, 094020  (2003).
%  doi:10.1103/PhysRevD.68.094020
%  [hep-ph/0306306].



%\cite{Ivanov:2006ni}
\bibitem{Ivanov:2006ni}
  M.A.~Ivanov, J.G.~K\"orner and P.~Santorelli,
%  ``Exclusive semileptonic and nonleptonic decays of the $B_c$ meson,''
  Phys.\ Rev.\ D {\bf 73}, 054024 (2006).
%  doi:10.1103/PhysRevD.73.054024
 % [hep-ph/0602050].



%\cite{Hernandez:2006gt}
\bibitem{Hernandez:2006gt}
E.~Hernandez, J.~Nieves, and J.M.~Verde-Velasco,
%``Study of exclusive semileptonic and non-leptonic decays of $B_c$ - in a nonrelativistic quark model,''
Phys. Rev. D \textbf{74}, 074008 (2006).
%doi:10.1103/PhysRevD.74.074008
%[arXiv:hep-ph/0607150 [hep-ph]].



%\cite{Naimuddin:2012dy}
\bibitem{Naimuddin:2012dy}
  S.~Naimuddin, S.~Kar, M.~Priyadarsini, N.~Barik, and P.~C.~Dash,
%  ``Nonleptonic two-body Bc-meson decays,''
  Phys.\ Rev.\ D {\bf 86}, 094028 (2012).
%  doi:10.1103/PhysRevD.86.094028



\bibitem{Kar:2013fna}
  S.~Kar, P.C.~Dash, M.~Priyadarsini, S.~Naimuddin, and N.~Barik,
  %``Nonleptonic $B_c \to VV$ decays,''
  Phys.\ Rev.\ D {\bf 88}, 094014 (2013).




%\cite{Ke:2013yka}
\bibitem{Ke:2013yka}
  H.W.~Ke, T.~Liu, and X.Q.~Li,
%  ``Transitions of $B_c\rightarrow \psi(1S,2S)$ and the modified
%    harmonic oscillator wave function in LFQM,''
  Phys.\ Rev.\ D {\bf 89},  017501 (2014).
%  doi:10.1103/PhysRevD.89.017501
%  [arXiv:1307.5925 [hep-ph]].



%\cite{Issadykov:2018myx}
\bibitem{Issadykov:2018myx}
A.~Issadykov and M.A.~Ivanov,
%``The decays $B_{c}\to J/\psi+\bar\ell\nu_\ell$ and $B_{c}\to J/\psi + \pi(K)$ in covariant confined quark model,''
Phys. Lett. B \textbf{783}, 178 (2018).
%doi:10.1016/j.physletb.2018.06.056
%[arXiv:1804.00472 [hep-ph]].


%\cite{Cheng:2021svx}
\bibitem{Cheng:2021svx}
W.~Cheng, Y.~Zhang, L.~Zeng, H.B.~Fu, and X.G.~Wu,
%``helicity form factors and the decays *,''
Chin. Phys. C \textbf{46}, 053103 (2022).
%doi:10.1088/1674-1137/ac4c9f
%[arXiv:2107.08405 [hep-ph]].



 %\cite{Xiao:2013lia}
\bibitem{Xiao:2013lia}
Z.J.~Xiao and X.~Liu,
%``The two-body hadronic decays of $B_c$ meson in the perturbative QCD approach: A short review,''
Chin. Sci. Bull. \textbf{59}, 3748 (2014).
%doi:10.1007/s11434-014-0418-z
%[arXiv:1401.0151 [hep-ph]].



 %\cite{Liu:2017cwl}
\bibitem{Liu:2017cwl}
X.~Liu, R.H.~Li, Z.T.~Zou and Z.J.~Xiao,
%``Nonleptonic charmless decays of $B_c\to TP, TV$ in the perturbative QCD approach,''
Phys. Rev. D \textbf{96}, 013005 (2017).
%doi:10.1103/PhysRevD.96.013005
%[arXiv:1703.05982 [hep-ph]].




\bibitem{Berger:2000wt}
  E.R.~Berger, A.~Donnachie, H.G.~Dosch, and O.~Nachtmann,
  %``Observing the odderon: Tensor meson photoproduction,''
  Eur.\ Phys.\ J.\ C {\bf 14}, 673 (2000).


\bibitem{Datta:2007yk}
  A.~Datta, Y.~Gao, A.V.~Gritsan, D.~London, M.~Nagashima, and A.~Szynkman,
  %``Study of Polarization in B ---> VT Decays,''
  Phys.\ Rev.\ D {\bf 77}, 114025 (2008).


\bibitem{Keum:2000ph}
  Y.Y.~Keum, H.-n.~Li and A.I.~Sanda,
  %``Fat penguins and imaginary penguins in perturbative QCD,''
  Phys.\ Lett.\ B {\bf 504}, 6 (2001);
%\bibitem{Keum:2000wi}
%  Y.Y.~Keum, H.-n.~Li and A.I.~Sanda,
  %``Penguin enhancement and B ---> K pi decays in perturbative QCD,''
  Phys.\ Rev.\ D {\bf 63}, 054008 (2001);
%\bibitem{Lu:2000em}
  C.D.~L\"u, K.~Ukai and M.Z.~Yang,
  %``Branching ratio and CP violation of B ---> pi pi decays in perturbative QCD approach,''
  Phys.\ Rev.\ D {\bf 63}, 074009 (2001);
%\bibitem{Li:2003yj}
  H.-n.~Li,
  %``QCD aspects of exclusive B meson decays,''
  Prog.\ Part.\ Nucl.\ Phys.\  {\bf 51}, 85 (2003).


\bibitem{Botts:1989kf}
  J.~Botts and G.F.~Sterman,
  %``Hard Elastic Scattering in QCD: Leading Behavior,''
  Nucl.\ Phys.\  {\bf B325}, 62 (1989);
%\bibitem{Li:1992nu}
  H.-n.~Li and G.F.~Sterman,
  %``The Perturbative pion form-factor with Sudakov suppression,''
  Nucl.\ Phys.\  {\bf B381}, 129 (1992).


\bibitem{Lu:2002ny}
  C.D.~L\"u and M.~Z.~Yang,
  %``B to light meson transition form-factors calculated in perturbative QCD approach,''
  Eur.\ Phys.\ J.\ C {\bf 28}, 515 (2003).



%\cite{Bondar:2004sv}
\bibitem{Bondar:2004sv}
A.E.~Bondar and V.L.~Chernyak,
%``Is the BELLE result for the cross section sigma(e+ e- ---> J / psi + eta(c)) a real difficulty for QCD?,''
Phys.\ Lett.\ B {\bf 612}, 215 (2005).


\bibitem{Chernyak:1983ej}
  V.L.~Chernyak and A.R.~Zhitnitsky,
  %``Asymptotic Behavior of Exclusive Processes in QCD,''
  Phys.\ Rep.\  {\bf 112}, 173 (1984);
%\bibitem{Zhitnitsky:1985dd}
A.R.~Zhitnitsky, I.R.~Zhitnitsky, and V.L.~Chernyak,
  %``Qcd Sum Rules And Properties Of Wave Functions Of Nonleading Twist. (in Russian),''
  Yad.\ Fiz.\  {\bf 41}, 445 (1985)
  [Sov.\ J.\ Nucl.\ Phys.\  {\bf 41}, 284 (1985)];
%\bibitem{Braun:1988qv}
  V.M.~Braun and I.E.~Filyanov,
  %``QCD Sum Rules in Exclusive Kinematics and Pion Wave Function,''
  Z.\ Phys.\ C {\bf 44}, 157 (1989); Yad.\ Fiz.\  {\bf 50}, 818 (1989)
  [Sov.\ J.\ Nucl.\ Phys.\  {\bf 50}, 511 (1989)];
%\bibitem{Braun:1989iv}
  V.M.~Braun and I.E.~Filyanov,
  %``Conformal Invariance and Pion Wave Functions of Nonleading Twist,''
  Z.\ Phys.\ C {\bf 48}, 239 (1990); Yad.\ Fiz.\  {\bf 52}, 199 (1990)
  [Sov.\ J.\ Nucl.\ Phys.\  {\bf 52}, 126 (1990)].


\bibitem{Ball:1998tj}
  P.~Ball,
  %``B ---> pi and B ---> K transitions from QCD sum rules on the light cone,''
  \jhep {\bf 09} (1998) 005;
%\bibitem{Ball:1998je}
%  P.~Ball,
  %``Theoretical update of pseudoscalar meson distribution amplitudes of higher twist: The Nonsinglet case,''
 % \jhep
 {\bf 01} (1999) 010.


\bibitem{Braun:2004vf}
  V.M.~Braun and A.~Lenz,
  %``On the SU(3) symmetry-breaking corrections to meson distribution amplitudes,''
  Phys.\ Rev.\ D {\bf 70}, 074020 (2004);
%\bibitem{Ball:2005ei}
  P.~Ball and A.N.~Talbot,
  %``Models for light-cone meson distribution amplitudes,''
  \jhep {\bf 06} (2005) 063;
%\bibitem{Ball:2005vx}
  P.~Ball and R.~Zwicky,
  %``SU(3) breaking of leading-twist K and K* distribution amplitudes: A Reprise,''
  Phys.\ Lett.\ B {\bf 633}, 289 (2006);
% \bibitem{Khodjamirian:2004ga}
  A.~Khodjamirian, T.~Mannel, and M.~Melcher,
  %``Kaon distribution amplitude from QCD sum rules,''
  Phys.\ Rev.\ D {\bf 70}, 094002 (2004).

%\cite{Cheng:2005nb}
\bibitem{Cheng:2005nb}
H.Y.~Cheng, C.K.~Chua and K.C.~Yang,
%``Charmless hadronic B decays involving scalar mesons: Implications to the nature of light scalar mesons,''
Phys. Rev. D \textbf{73}, 014017 (2006);
%\cite{Cheng:2007st}
%\bibitem{Cheng:2007st}
%H.~Y.~Cheng, C.~K.~Chua and K.~C.~Yang,
%``Charmless B decays to a scalar meson and a vector meson,''
%Phys. Rev. D
\textbf{77}, 014034 (2008).

%\cite{Li:2008tk}
\bibitem{Li:2008tk}
R.H.~Li, C.D.~L\"u, W.~Wang and X.X.~Wang,
%``B ---\ensuremath{>} S Transition Form Factors in the PQCD approach,''
Phys. Rev. D \textbf{79}, 014013 (2009).


\bibitem{Ball:vector}
P.~Ball, V.M.~Braun, Y.~Koike and K.~Tanaka,  Nucl.\ Phys.\  {\bf B529}, 323 (1998);
% [hep-ph/9802299];
 P.~Ball and V.M.~Braun,  Nucl.\ Phys.\  {\bf B543}, 201 (1999);
 % [hep-ph/9810475].
%\bibitem {previousvectorwf}
%P.~Ball and  V.M.~Braun,
\prd {\bf 54}, 2182 (1996);
%[hep-ph/9602323];
 P.~Ball and R.~Zwicky,\prd {\bf 71}, 014029 (2005);
 \jhep {\bf 02} (2006) 034;
% \jhep
 {\bf 04} (2006) 046;
 % [hep-ph/0601086];
P.~Ball and M.~Boglione, \prd {\bf
68}, 094006 (2003).
% [hep-ph/0307337].


\bibitem{Ball:2007rt}
  P.~Ball and G.W.~Jones,
  %``Twist-3 distribution amplitudes of K* and phi mesons,''
  \jhep {\bf 03} (2007) 069.


\bibitem{Yang:2007zt}
  K.C.~Yang,
  %``Light-cone distribution amplitudes of axial-vector mesons,''
  Nucl.\ Phys.\  {\bf B776}, 187 (2007);
%\bibitem{Yang07:twist}
%K.C.~Yang,
\jhep {\bf 10} (2005) 108;
% Light-cone distribution amplitudes for the light 1$^1P_1$ mesons
%  \npb {\bf 776}, 187 (2007).
%``Light-cone distribution amplitudes of axial-vector mesons,''
%\cite{Cheng:2008gxa}
%\bibitem{Cheng:2008gxa}
H.Y.~Cheng and K.C.~Yang,
%``Branching Ratios and Polarization in B ---\ensuremath{>} VV, VA, AA Decays,''
Phys. Rev. D \textbf{78}, 094001 (2008);
%[Erratum: Phys. Rev. D
\textbf{79}, 039903(E) (2009).


\bibitem{Li:2009tx}
  R.H.~Li, C.D.~L\"u, and W.~Wang,
  %``Transition form factors of B decays into p-wave axial-vector mesons in the perturbative QCD approach,''
  Phys.\ Rev.\ D {\bf 79}, 034014 (2009).




\bibitem{Cheng:2010hn}
  H.Y.~Cheng, Y.~Koike, and K.C.~Yang,
  %``Two-parton Light-cone Distribution Amplitudes of Tensor Mesons,''
  Phys.\ Rev.\ D {\bf 82}, 054019 (2010).



\bibitem{Wang:2010ni}
  W.~Wang,
  %``B to tensor meson form factors in the perturbative QCD approach,''
  Phys.\ Rev.\ D {\bf 83}, 014008 (2011).


\bibitem{Buchalla:1995vs}
  G.~Buchalla, A.J.~Buras, and M.E.~Lautenbacher,
  %``Weak decays beyond leading logarithms,''
  Rev.\ Mod.\ Phys.\  {\bf 68}, 1125 (1996).


\bibitem{Zou:2015iwa}
  Z.T.~Zou, A.~Ali, C.D.~L\"u, X.~Liu, and Y.~Li,
  %``Improved Estimates of The $B_{(s)}\to V V$ Decays in Perturbative QCD Approach,''
  Phys.\ Rev.\ D {\bf 91}, 054033 (2015).



%\cite{Fleischer:2011au}
\bibitem{Fleischer:2011au}
R.~Fleischer, R.~Knegjens and G.~Ricciardi,
%``Anatomy of $B^0_{s,d} \to J/\psi f_0(980)$,''
Eur. Phys. J. C \textbf{71}, 1832 (2011).


\bibitem{Wolfenstein:1983yz}
  L.~Wolfenstein,
  %``Parametrization of the Kobayashi-Maskawa Matrix,''
  Phys.\ Rev.\ Lett.\  {\bf 51}, 1945 (1983).


%\cite{Sun:2007ei}
\bibitem{Sun:2007ei}
J.f.~Sun, G.f.~Xue, Y.l.~Yang, G.~Lu, and D.s.~Du,
%``Study of $B_c \to J/psi\pi^-, \eta_c\pi^-$ decays with QCD factorization,''
Phys. Rev. D \textbf{77}, 074013 (2008).
%doi:10.1103/PhysRevD.77.074013
%[arXiv:0710.0031 [hep-ph]].


%\cite{Du:2022nno}
\bibitem{Du:2022nno}
M.C.~Du, Y.~Cheng, and Q.~Zhao,
%``Vertex corrections due to the triangle singularity mechanism in the light axial vector meson couplings to K*K\textasciimacron{}+c.c.,''
Phys. Rev. D \textbf{106}, 054019 (2022).


\bibitem{Cheng:2011pb}
  H.Y.~Cheng,
  %``Revisiting Axial-Vector Meson Mixing,''
  Phys.\ Lett.\ B {\bf 707}, 116 (2012).


%\cite{Liu:2014dxa}
\bibitem{Liu:2014dxa}
X.~Liu, Z.T.~Zou and Z.J.~Xiao,
%``Penguin-dominated B\textrightarrow{}\ensuremath{\phi}K$_1$(1270) and \ensuremath{\phi}K$_1$(1400) decays in the perturbative QCD approach,''
Phys. Rev. D \textbf{90}, 094019 (2014).



\bibitem{Aubert:2007kpb}
  B.~Aubert {\it et al.} ({\it BABAR} Collaboration),
  %``Evidence for charged B meson decays to a+-(1)(1260) pi0 and a0(1)(1260) pi+-,''
  Phys.\ Rev.\ Lett.\  {\bf 99}, 261801 (2007).



%\cite{Ali:1978kn}
\bibitem{Ali:1978kn}
A.~Ali, J.G.~Korner, G.~Kramer, and J.~Willrodt,
%``Nonleptonic Weak Decays of Bottom Mesons,''
Z. Phys. C \textbf{1}, 269 (1979).

%\cite{Suzuki:2001za}
\bibitem{Suzuki:2001za}
M.~Suzuki,
%``Final-state interactions and $s^-$ quark helicity conservation in $B \to J/\psi K^{*}$,''
Phys. Rev. D \textbf{64}, 117503 (2001).


\bibitem{Wang:2017bgv}
  W.~Wang, J.~Xu, D.~Yang, and S.~Zhao,
  %``Relativistic corrections to light-cone distribution amplitudes of S-wave B$_{c}$ mesons and heavy quarkonia,''
  \jhep {\bf 12} (2017) 012.

%\cite{Sun:2015exa}
\bibitem{Sun:2015exa}
J.~Sun, N.~Wang, Q.~Chang, and Y.~Yang,
%``$B_{c}$ ${\to}$ $BP$, $BV$ decays with the QCD factorization approach,''
Adv.\ High Energy Phys.\  {\bf 2015}, 104378 (2015).




\bibitem{Chiu:2007km}
  T.W.~Chiu {\it et al.} (TWQCD Collaboration),
  %``Beauty mesons in lattice QCD with exact chiral symmetry,''
  Phys.\ Lett.\ B {\bf 651}, 171 (2007).



\bibitem{Li05:kphi}
H.-n.~Li, and S.~Mishima, \prd {\bf 71}, 054025 (2005);
% Polarizations in B \to VV decays
H.-n.~Li, \plb {\bf 622}, 63 (2005).
% Resolution to the $B \to  \phi K^*$ polarization puzzle



%\cite{Chen:2021dwn}
\bibitem{Chen:2021dwn}
Y.~Chen, Z.~Jiang, and X.~Liu,
%``Pure annihilation decays of \$B\textasciicircum{}0\_S \textbackslash{}to a\_0\textasciicircum{}+
%a\_0\textasciicircum{}-$ and $B\textasciicircum{}0\_d \textbackslash{}to K\textasciicircum{}{*+}\_0 K\textasciicircum{}{*-}\_0\$ in the PQCD approach,''
Commun. Theor. Phys. \textbf{73}, 045201 (2021).

\bibitem{Lu:2006fr}
  C.D.~L\"u, Y.M.~Wang, and H.~Zou,
  %``Twist-3 distribution amplitudes of scalar mesons from QCD sum rules,''
  Phys.\ Rev.\ D {\bf 75}, 056001 (2007).



\end{thebibliography}
\end{document}